\documentclass[10pt,letterpaper]{article}
\usepackage[top=1in,left=1in,footskip=0.75in]{geometry}
\usepackage[utf8x]{inputenc}
\usepackage{amsmath,amssymb}
\usepackage{color,soul}
\usepackage{enumitem}
\usepackage{cite}
\usepackage{nameref,hyperref}
\usepackage[table]{xcolor}
\usepackage{tcolorbox}
\definecolor{arsenic}{rgb}{0.23, 0.27, 0.29}
\definecolor{bluegray}{rgb}{0.4, 0.6, 0.8}
\definecolor{burgundy}{rgb}{0.5, 0.0, 0.13}
\usepackage{float}
\usepackage{textgreek}
\usepackage[right]{lineno} 
\usepackage[aboveskip=1pt,labelfont=bf,labelsep=period,justification=raggedright,singlelinecheck=off]{caption}

\usepackage{lastpage,fancyhdr,graphicx}
\usepackage{epstopdf}
\usepackage{marvosym}

\usepackage{placeins}

\begin{document}
\begin{flushleft}
{\Large
\textbf\newline{A single-cell mathematical model of SARS-CoV-2 induced pyroptosis and the effects of anti-inflammatory intervention
} 
}

\bigskip
Sara J Hamis,\textsuperscript{1\Yinyang}
Fiona R Macfarlane.\textsuperscript{1\Yinyang}
\\
\bigskip
\textbf{1} School of Mathematics and Statistics, University of St Andrews, St Andrews, Scotland, UK.
\\
\bigskip
\Yinyang \: Both authors contributed equally to this work.
\\
\bigskip
{\it This is version \textbf{4} of the pre-print article.}  \end{flushleft}

\vspace{0.35cm}

\begin{abstract} 
\noindent Pyroptosis is an inflammatory mode of cell death that can contribute to the cytokine storm associated with severe cases of coronavirus disease 2019 (COVID-19). The formation of the NLRP3 inflammasome is central to pyroptosis, which may be induced by severe acute respiratory syndrome coronavirus 2 (SARS-CoV-2). Inflammasome formation, and by extension pyroptosis, may be inhibited by certain anti-inflammatory drugs. 

In this study, we present a single-cell mathematical model that captures the formation of the NLRP3 inflammasome, pyroptotic cell death and responses to anti-inflammatory intervention that hinder the formation of the NLRP3 inflammasome. The model is formulated in terms of a system of ordinary differential equations (ODEs) that describe the dynamics of the key biological components involved in pyroptosis. Our results demonstrate that an anti-inflammatory drug can delay the formation of the NLRP3 inflammasome, and thus may alter the mode of cell death from inflammatory (pyroptosis) to non-inflammatory ({\em e.g.}, apoptosis).
The single-cell model is implemented within a SARS-CoV-2 tissue simulator, in collaboration with a multidisciplinary coalition investigating within host-dynamics of COVID-19. 
In this paper, we additionally provide an overview of the SARS-CoV-2 tissue simulator and highlight the effects of pyroptosis on a cellular level.
\end{abstract}


\section{Introduction}
\subsection{SARS-CoV-2 induced pyroptosis and cytokine storms}
COVID-19 is a respiratory illness induced by the coronavirus strain SARS-CoV-2~\cite{zhou2020pneumonia}.
Most people infected by the virus experience mild symptoms, however, in severe cases of the disease, life-threatening symptoms can manifest~\cite{Yap2020}.
These divergent disease trajectories have been, largely, attributed to differences in the immune response of infected hosts~\cite{Tay2020}.
When cells register the presence of SARS-CoV-2 virions, a multitude of host-protective responses are triggered.
Infected cells transmit signals that recruit immune cells, such as monocytes, macrophages and T-cells, to the site of infection.
These immune cells act to eliminate the virus from the body, and thus they attack infected cells in which virions may replicate.
Current research has shown that, upon active virion replication and release, SARS-CoV-2 can induce pyroptosis in both epithelial cells and immune cells~\cite{Tay2020,Soy2020, Tang2020,shah2020novel,ratajczak2020sars,fu2020understanding,nagashima2020endothelial}.

Pyroptosis is an inflammatory and rapid mode of cell death that is characterised by the secretion of pro-inflammatory cytokines, cell swelling and, ultimately, membrane rupture resulting in the release of cytoplasmic contents into the extracellular environment~\cite{Jamilloux2020,Taabazuing2017}. 
Furthermore, interleukin-1$\beta$ (IL-1$\beta$), a cytokine released by pyroptosing cells, has been shown to induce pyroptosis in neighbouring bystander cells~\cite{kelley2019nlrp3}. The cytokine interleukin-18 (IL-18), which is also released by pyroptosing cells, can act as a recruiter of immune cells~\cite{Bergsbaken2009,dinarello2009immunological,he2016mechanism,zalinger2017role,Stutz2009,oshea2015cytokines}. 
Fundamentally, both the recruitment of immune cells and pyroptotic cell death act to protect the host from the virus. 
In hosts with healthy immune systems, virus-specific T-cells are recruited to the site of infection~\cite{Tay2020}, and cytokine levels are kept under control by negative-feedback regulations~\cite{Song2020}. 
However, if the immune system contesting a viral infection is malfunctioning, a wide array of cytokines may be over-produced due to deregulation of the negative-feedback that controls cytokine levels in healthy immune systems~\cite{Song2020}.
Such an over-compensating immune response may lead to uncontrolled cytokine activity, commonly referred to as a {\it cytokine storm}~\cite{Tay2020}.
Elevated levels of both pro-inflammatory and anti-inflammatory cytokines have been observed in severe cases of COVID-19~\cite{Yap2020,Song2020,huang2020clinical,nagashima2020endothelial}, commonly manifesting with severe symptoms such as pneumonia and acute respiratory distress syndrome (ARDS), which may lead to multiple organ failure~\cite{Song2020}.
Hence, cytokine storms are associated with poor clinical COVID-19 outcomes~\cite{Soy2020, Yap2020,shah2020novel}.
Therefore, suppressing the onset of cytokine storms and the subsequent inflammation, without completely cancelling-out host-protective effects of the immune system, is one of the suggested treatment strategies being explored to combat symptoms of COVID-19~\cite{Song2020}. 

One approach to suppress cytokine storms is to inhibit pyroptosis, whilst leaving 
other functionalities of the host immune response untouched.
In fact, there exists a number of anti-inflammatory drugs that may inhibit pyroptosis and, through this inhibition of pyroptosis, alter the mode of cell death from inflammatory to non-inflammatory~\cite{Bertinaria2019}.
Non-inflammatory modes of cell death include apoptosis, which is characterised by cell shrinkage with cell membrane integrity maintained throughout cell death, whereas pyroptosis is characterised by cell swelling and membrane rupture~\cite{Tsuchiya2020,Fink2005}.
Inhibiting pyroptosis may provide two key clinical advantages. Firstly, this inhibition could suppress the onset and effects of cytokine storms, and the resulting increased inflammation and tissue damage that they generate. Secondly, it has been shown that tissue factors released upon pyroptosis may initiate blood coagulation cascades, therefore inhibiting pyroptosis may reduce the risk of blood clotting in COVID-19 cases~\cite{Yap2020}.

\FloatBarrier

\subsection{The pathway to pyroptosis}
In this subsection we describe the key aspects of the intracellular pathway that leads to pyroptosis. 
Using biological details from this well-established pathway, we formulate a mathematical model in Section~\ref{sect:model} that describes pyroptotic cell death. The model components are illustrated in Figure~\ref{fig:intro_pathway}. 
For a more comprehensive description of the mechanisms driving pyroptosis, we refer the interested reader to detailed reviews~\cite{Kozloski2020,Stutz2009,Christgen2020,Bertinaria2019,Zheng2020}. 

{\begin{figure}
\begin{center}
 \includegraphics[width=\textwidth]{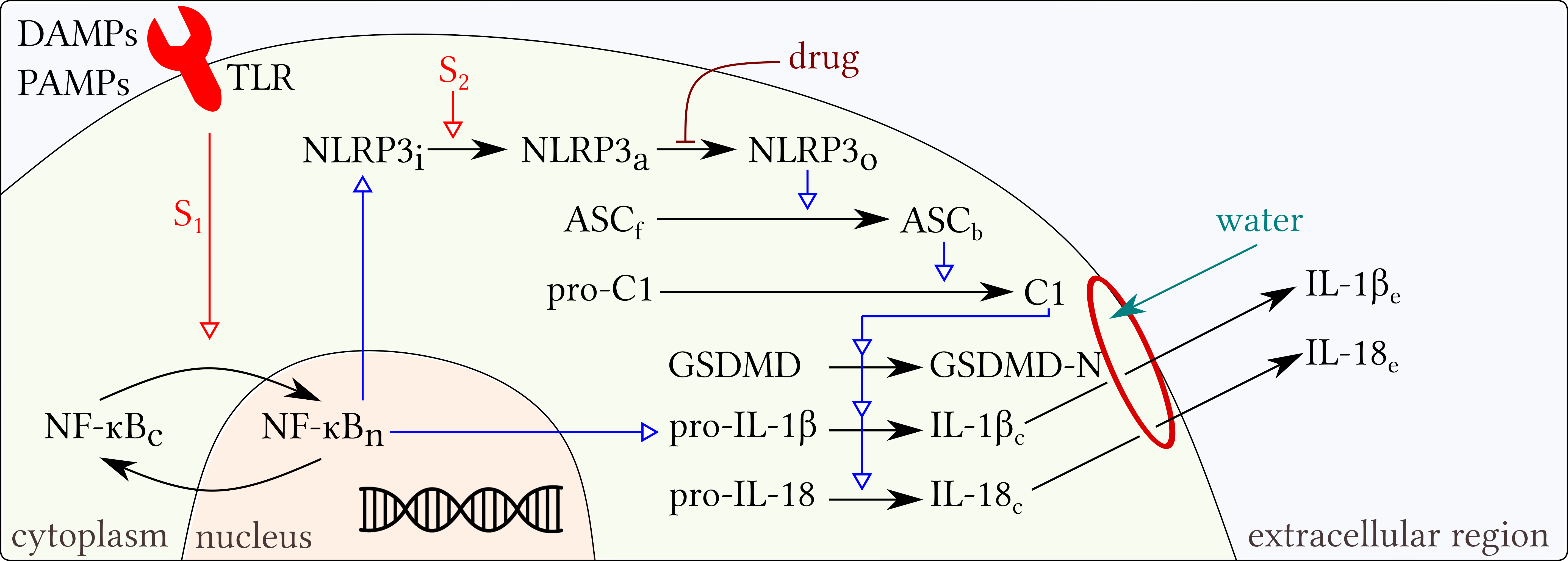}
 \caption{A schematic representation of the pathway to pyroptosis, as considered in our mathematical model. Black arrows represent reactions that involve mass transfer between depicted model components. Blue arrows represent facilitation of component formation or reactions. Red arrows represent signals that can be turned {\it on} or {\it off} in the model. GSDMD-N induced membrane pores (shown in red) allow for the outflux of inflammatory cytokines and the influx of extracellular water, causing the cell to swell and the cell membrane to eventually rupture. Our mathematical model includes the effect of a drug that inhibits the NLRP3\textsubscript{o} inflammasome base from forming. Figure abbreviations are listed in Appendix~\ref{appendix_abb_intext}. 
 }
 \label{fig:intro_pathway}
 \end{center}
\end{figure}}

The formation of the NLRP3 inflammasome is a fundamental step in the pyroptosis pathway. This multi-protein complex consists of three molecular units (a sensor, an adaptor and an executor) and enables the cleavage the executor molecule caspase-1, which regulates pore formation and the release of cytokines in pyroptosis.
Note that, only one inflammasome is formed per cell~\cite{Stutz2009}.
The inflammasome is named after its sensor molecule NLRP3, which is short for nucleotide-binding and oligomerisation domain (NBD) leucine-rich repeat (LRR)-containing receptors with an N-terminal pyrin domain (PYD) 3. 
In this study we focus our attention on the inflammasome comprising the sensor molecule NLRP3, the adaptor molecule apoptosis-associated speck-like protein (ASC), and caspase-1. 

Homeostatic cells do not contain enough NLRP3 to produce an inflammasome base~\cite{Christgen2020}. Instead, NLRP3 levels are elevated in cells when toll like receptors (TLRs) on the cell surface sense damage associated molecular patterns (DAMPs) or pathogen associated molecular patterns (PAMPs).
As SARS-CoV-2 is a positive-sense RNA virus, it can be detected by TLRs and induce NLRP3 inflammasome formation~\cite{Yap2020,shah2020novel,ratajczak2020sars}.
Upon TLRs detecting DAMPs or PAMPs, the transcription factor NF-$\kappa$B is translocated to the nucleus, initiating the transcription and, by extension, the synthesis of (inactive) NLRP3. The transcription and synthesis of the pro-inflammatory cytokine pro-interleukin-1$\beta$ (pro-IL-1$\beta$) are also regulated by NF-$\kappa$B.
The increased transcription of (inactive) NLRP3 and pro-IL-1$\beta$ is often referred to as the {\it priming} step of pyroptosis, which is succeeded by the {\it activation} step~\cite{Stutz2009}.

\begin{figure}
\begin{center}
\includegraphics[width=\textwidth]{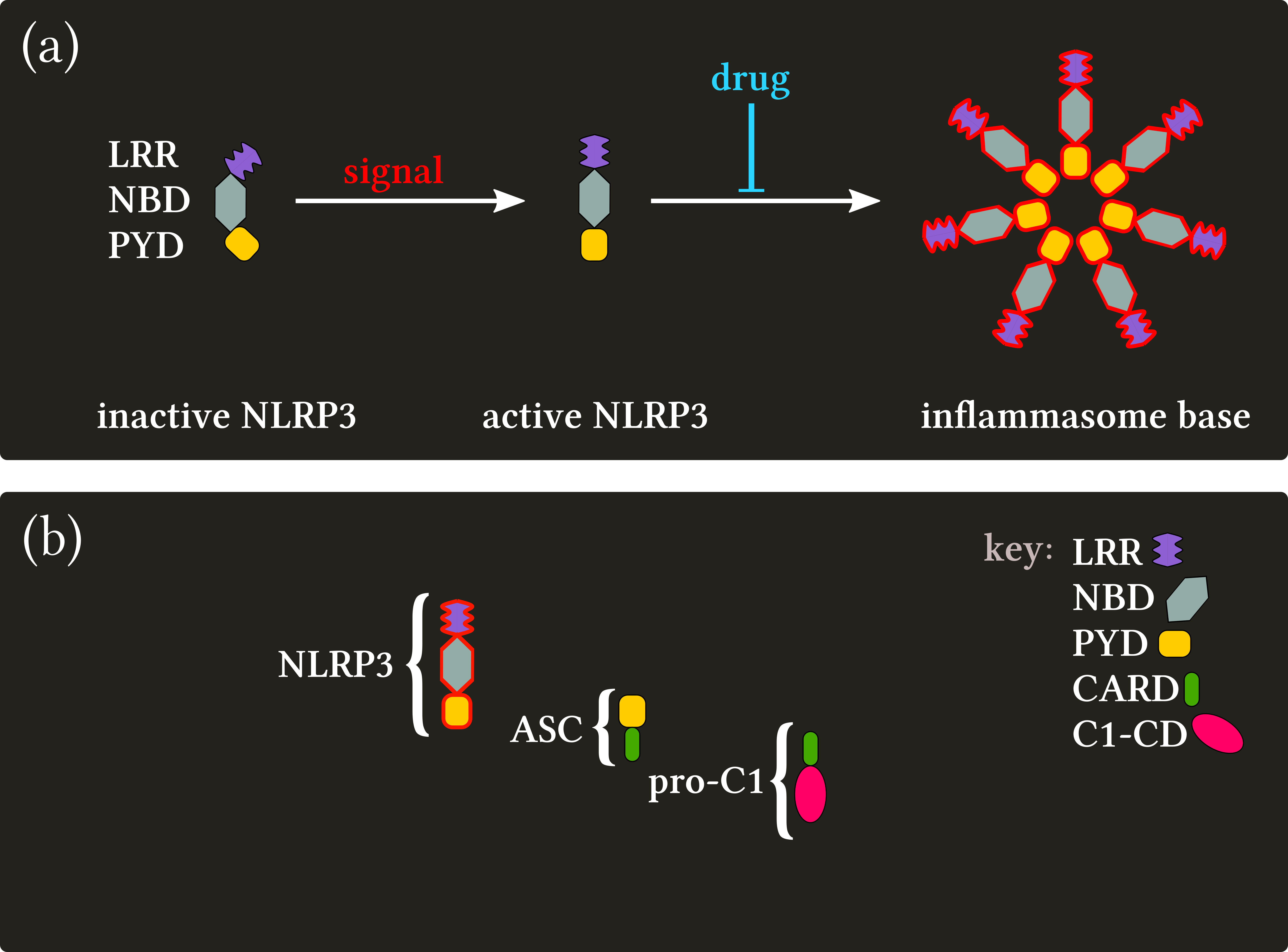}
\caption{NLRP3 inflammasome formation. (a) Through a wide array of stimuli, inactive NLRP3 can become activated. Active NLRP3 can bind together to form the inflammasome base, this binding can be inhibited by NLRP3-targeting anti-inflammatory drugs. 
(b) Once the inflammasome base is formed, the pyrin domain (PYD) of ASC molecules can bind to the PYD of NLRP3 molecules in the inflammasome base. Thereafter, the caspase activation and recruitment domain (CARD) of pro-caspase-1 can bind to the CARD of NLRP3-bound ASC molecules. This results in the dimerisation and activation of caspase-1. Abbreviations used in the figure are listed in Appendix~\ref{appendix_abb_intext}. 
 }
 \label{fig:intro_inflammasome}
 \end{center}
\end{figure}

Inactive NLRP3 can become activated upon stimuli from a wide array intracellular events, including altered calcium signalling, potassium efflux and the generation of reactive oxygen species (ROS)~\cite{Bertinaria2019}. 
In turn, active NLRP3 molecules can oligomerise, {\em i.e.,} bind together, to form a wheel shaped structure, constituting the inflammasome base as depicted in Figure~\ref{fig:intro_inflammasome}(a).
Through homotypic binding, ASC molecules can then bind to the NLRP3 inflammasome base as illustrated in Figure~\ref{fig:intro_inflammasome}(b).
Thereafter, pro-caspase-1 can bind to ASC within the inflammasome, enabling the proximity-induced dimerisation, and thereby the activation, of caspase-1~\cite{Sanders2015}.
Active caspase-1 then mediates the cleavage of the pro-interleukins, pro-IL-1$\beta$ and pro-IL-18, into their respective mature forms, IL-1$\beta$ and IL-18.
Caspase-1 also cleaves the protein gasdermin D (GSDMD), releasing the active N-terminal domain of gasdermin D (GSDMD-N) which can form pores on the cell membrane.
These pores enable the {\it outflux} of the inflammatory cytokines IL-1$\beta$ and IL-18 (in their mature forms) from the cytoplasm to the external environment~\cite{semino2018progressive,lopez2011understanding}.
The released cytokines can recruit immune cells to the site of infection and initiate pyroptosis in neighbouring cells.
Furthermore, the GDSMD-N derived membrane pores also allow for the {\it influx} of extracellular material, {\em e.g.}, water, into the cell.
This influx causes cells to swell until their plasma membrane eventually ruptures, releasing cellular material into the extracellular region.

Data suggests that ASC, pro-caspase-1, GSDMD and pro-IL-18 are present in adequate levels for NLRP3 formation and pyroptosis at homeostasis, and thus these proteins are not upregulated by the TLR-stimulated cytoplasm-to-nucleus translocation of NF-$\kappa$B~\cite{Kozloski2020,broz2019gasdermins}.
Furthermore, once the NLRP3 inflammasome is activated, cell lysis can be delayed but not prevented~\cite{dipeso2017cell,brough2007caspase}.
However, if the formation of the NLRP3 inflammasome is delayed, a series of intracellular events resulting in non-inflammatory cell death may instead be initiated. There exists multiple drugs that act to inhibit inflammasome formation in order to prevent pyroptosis, a rigorous list of covalent drugs that target the NLRP3 inflammasome can be found in a review by Bertinara {\em et al.}~\cite{Bertinaria2019}.
In this study, we include the pharmacodynamical effects of a generic anti-inflammatory drug which inhibits the formation of the NLRP3 inflammasome by covalently binding to NLRP3 molecules to prevent NLRP3-NLRP3 interactions, as shown in Figure~\ref{fig:intro_inflammasome}(a).

\subsection{Mathematical modelling of pyroptosis}
The use of mathematical models to describe apoptotic cell death has been well established and reviewed~\cite{schleich2013mathematical,spencer2011measuring}. However, there are significantly fewer mathematical models that describe pyroptotic cell death. Previous modelling works include more implicit descriptions of pyroptosis~\cite{wang2018caspase} and descriptions of specific aspects of inflammasome formation~\cite{veltman2017signal,bozkurt2015unified,lopez2018stable}. Within this work, we explicitly model the intracellular events that drive SARS-CoV-2 induced pyroptosis, from TLRs detecting DAMPs/PAMPs through to the ultimate membrane rupture. We study this system in the absence, or presence, of a generic anti-inflammatory drug. The goal of our model is to capture the key aspects of the pyroptosis pathway that are commonly analysed in experimental studies. Thus, in this mathematical/computational work, we study the temporal evolution of NF-$\kappa$B, the NLRP3 inflammasome and its components, the pore-forming protein GSDMD, the pro-inflammatory cytokines IL-1$\beta$ and IL-18, and the cell volume. 

\subsection{Mathematical modelling of within host-dynamics of COVID-19}
Since COVID-19 was announced as a global pandemic~\cite{WHO}, many researchers have developed and used mathematical tools and models to study the SARS-CoV-2 virus and COVID-19. 
These models can be split into two main categories, which are, external dynamics models (\emph{e.g.}, models that consider the spread of SARS-CoV-2 from person to person) and within-host models (\emph{e.g.}, models that capture the dynamics of SARS-CoV-2 and SARS-CoV-2 induced responses within the body). 
Agent-based models (ABMs) have been widely utilised in the past to model within-host viral dynamics~\cite{akpinar2016spatial,bankhead2013simulation,bauer2009agent,bocharov2016spatiotemporal}, with focuses on influenza~\cite{beauchemin2006modeling,beauchemin2006probing,beauchemin2005simple,levin2016spatial,fachada2009simulating}, oncolytic virotherapy~\cite{jenner2020enhancing,wodarz2012complex} and inflammation~\cite{an2008introduction,cockrell2018examining}. 
ABMs can be used to describe how viruses spread within the body and circumvent the human immune system. Accordingly, several research groups are currently using multiscale ABMs to describe the within-host dynamics in response to SARS-CoV-2~\cite{fatehi2020comparing,sego2020modular,wang2020rapid}. As we describe in Section \ref{sect:abm}, our model of single-cell pyroptotic cell death is incorporated within a multiscale SARS-CoV-2 tissue simulator, that has been developed in response to the COVID-19 pandemic by an international and interdisciplinary research coalition. 
For the full details of this tissue simulator, we refer the reader to the project preprint~\cite{wang2020rapid}. 

\section{Mathematical model of pyroptosis in a single-cell}
\label{sect:model}
We formulate a system of ordinary differential equations (ODEs) describing the dynamics of the key components of NLRP3 inflammasome formation and pyroptosis. Specifically, the model includes the dynamics of NF-$\kappa$B (Section~\ref{sec:model_nfkb}), NLRP3 (Section~\ref{sec:model_nlrp3}), ASC (Section~\ref{sec:model_asc}), caspase-1 (Section~\ref{sec:model_caspase1}), GSDMD (Section~\ref{sec:model_gsdmd}), IL-1$\beta$ and IL-18 (Section~\ref{sec:model_il1b8}) and cell volume. 
Figure~\ref{fig:intro_pathway} provides a pictorial overview of the modelled pathway, and variable names are listed in Table~\ref{tab:var}. Throughout this work, the bracket notation [ ] denotes molecule concentrations. Further details concerning modelling choices, reaction terms and parameterisation can be found in Appendices~\ref{app_description} and~\ref{app_parameters}. 
In Section~\ref{sect:drug}, we expand the model to include the pharmacodynamic effects of an anti-inflammatory drug targeting NLRP3. We include the full system of equations and describe the set-up of the numerical simulations in Section~\ref{sect:implement}. The results of these numerical simulations are described in Section~\ref{sect:results}, and we further investigate the model using sensitivity analysis in Section \ref{sect:sens}.

\subsection{NF-$\kappa$B dynamics}
\label{sec:model_nfkb}
When TLRs register DAMPs or PAMPs, cytoplasmic NF-$\kappa$B is translocated to the nucleus in order to initiate the transcription of inactive NLRP3 and pro-IL-1$\beta$. In the case of SARS-CoV-2, these DAMPs and PAMPs can be induced by intracellular virion replication and release, or by cytokines from neighbouring, infected cells. 
Dynamics of NF-$\kappa$B translocation between the cytoplasm and the nucleus vary between situations and strongly depend on cell type and external stimuli. 
In recent work, Bagaev {\em et al.} \cite{bagaev2019elevated} studied TLR-4 derived activation of NF-$\kappa$B in bone marrow derived macrophages (BMDMs) subjected to bacterial lipopolysaccharide (LPS). 
The authors found a sharp peak in NF-$\kappa$B nuclear translocation kinetics 10 minutes post LPS activation, after which the nuclear NF-$\kappa$B signalling gradually decreased until plateau. 
{BMDMs stimulated with LPS have been shown to undergo pyroptosis \cite{ZhangZhaoChen_2020, Huang2018}, and we therefore aim to formulate a mathematical model that achieves nuclear NF-$\kappa$B levels that, over time, are $(i)$ initially rapidly increasing, $(ii)$ mono-peaking and $(iii)$ right-skewed. 
In order to achieve $(i-iii)$, we here choose to phenomenologically model nuclear NF-$\kappa$B levels using the function, 
\begin{equation}
\label{eq:expectef_nfkbn}
\text{[NF-$\kappa$B\textsubscript{n}]}({t}) = \text{[NF-$\kappa$B\textsubscript{n}]}(0) + S_1 \cdot h\ e^{- \frac{\log^2 (t/\tau)}{s} },
\end{equation}
where [NF-$\kappa$B\textsubscript{n}](0) is the amount of the cell's NF-$\kappa$B that is located in the nucleus prior to TLR-signalling. The DAMP/PAMP initiated TLR-signal, $S_1$, has been explicitly included as a binary {\it on/off} function such that,
\begin{equation}
 S_{1}:= \begin{cases} 1 &\mbox{{\it on}, if a DAMP/PAMP induced signal is present, }\\
0 & \mbox{{\it off,} otherwise.} \end{cases} \label{eqS1}
\end{equation}
In Equation~\eqref{eq:expectef_nfkbn}, $h$ is a constant that denotes the peak elevation of the [NF-$\kappa$B\textsubscript{n}] value, and 
$\tau$ denotes the time point (in units of minutes) at which the peak occurs. The variable $s$ controls the skewness of the [NK-$\kappa$B\textsubscript{n}] peak over time. 
Accordingly, the parameters $h$, $\tau$ and $s$ can be varied to adjust the shape of the [NF-$\kappa$B\textsubscript{n}] peak, as is shown in the Supplementary Material (Supplementary Material, S1). 
}

\subsection{NLRP3 dynamics}
\label{sec:model_nlrp3}
Following nuclear NF-$\kappa$B translocation, inactive NLRP3, NLRP3\textsubscript{i}, is transcribed and subsequently synthesised. As the transcription and synthesis of inactive NLRP3 is promoted by nuclear NF-$\kappa$B, we describe the production rate of NLRP3 using a Hill function~\cite{Salahudeen_ligandbinding}. Specifically, a Hill function with a constant coefficient rate $\alpha_{1}$, Hill coefficient $\gamma_{\text{NF}}$ and half-max concentration-value NF\textsubscript{50}. Inactive NLRP3 can become activated in response to a secondary signal, $S_{2}$, which can be turned {\it on} by multiple stimuli such as potassium influx, calcium outflux and abnormal reactive oxygen species (ROS). Once activated, NLRP3\textsubscript{a} molecules can oligomerise (bind together) to form a wheel-shaped inflammasome base, modelled by the concentration of oligomerised NLRP3, {\em i.e.,} NLRP3\textsubscript{o}. In the model, we thus consider NLRP3 protein concentrations in three different forms: [NLRP3\textsubscript{i}], [NLRP3\textsubscript{a}] and [NLRP3\textsubscript{o}]. 
Inactive and active NLRP3 decays in the model at a rate $\delta_{1}$, however we do not include the decay of NLRP3\textsubscript{o} in the model. The reason behind this choice of omitting NLRP3\textsubscript{o} decay is that, in experimental works, the functions of the NLRP3 inflammasome are evident until cell death in the context of pyroptosis~\cite{cheng2010kinetic,han2015lipopolysaccharide,deVasconcelos2019}. Furthermore, it is unclear whether disassembly of the inflammasome is possible in any context~\cite{chai2014apoptosome,ruland2014inflammasome}. The forward reactions from the inactive-to-active, and active-to-oligomerised NLRP3 forms are here assumed to be irreversible, and are respectively described using the rates $k_{1}$ and $k_{2}$.

We now incorporate all the above mechanisms to describe the rate of change of [NLRP3\textsubscript{i}], [NLRP3\textsubscript{a}] and [NLRP3\textsubscript{o}] over time as,
\begin{eqnarray}
\frac{d\text{[NLRP3\textsubscript{i}]}}{dt}&=& \alpha_{1} \frac{[\overline{\text{NF-$\kappa$B\textsubscript{n}}}]^{\gamma_{\text{NF}}}}{\text{NF}_{50}^{\gamma_{\text{NF}}}+[\overline{\text{NF-$\kappa$B\textsubscript{n}}}]^{\gamma_{\text{NF}}}}-S_{2}\ k_{1}\ \text{[NLRP3\textsubscript{i}]} -\delta_{1}\ \text{[NLRP3\textsubscript{i}]}, \label{eqNi}\\
\frac{d\text{[NLRP3\textsubscript{a}]}}{dt}&=&S_{2}\ k_{1}\ \text{[NLRP3\textsubscript{i}]}-k_{2}\ \text{[NLRP3\textsubscript{a}]}^2-\delta_{1}\ \text{[NLRP3\textsubscript{a}]}, \label{eqNa}\\
\frac{d\text{[NLRP3\textsubscript{o}]}}{dt}&=&k_{2}\ \text{[NLRP3\textsubscript{a}]}^2,\label{eqNo}
\end{eqnarray}
where [$\overline{\text{NF-$\kappa$B\textsubscript{n}}}$] denotes the deviation from baseline in [NF-$\kappa$B\textsubscript{n}] so that, 
\begin{equation}
\label{eq:nfkb_shift_for_transcription}
 [\overline{\text{NF-$\kappa$B\textsubscript{n}}}](t) = \text{[NF-$\kappa$B\textsubscript{n}]}(t) - \text{[NF-$\kappa$B\textsubscript{n}]}(0). 
\end{equation}
The binary signal $S_2$ is set as,
\begin{equation}
 S_{2}:= \begin{cases} 1 &\mbox{{\it on}, if an activation signal is present, }\\
0 & \mbox{{\it off,} otherwise.} \end{cases} \label{eqS2}
\end{equation}

\subsection{ASC dynamics}
\label{sec:model_asc}
Once the inflammasome base is formed, free ASC in the cell can bind to the NLRP3 inflammasome base. Thus in the model, we consider the concentrations of ASC in free and bound form, denoted [ASC\textsubscript{f}] and [ASC\textsubscript{b}], respectively. 
We consider this binding process to be irreversible and to occur at a rate $k_{3}$, hence the rate of change of [ASC\textsubscript{f}] and [ASC\textsubscript{b}] are here described by,
\begin{eqnarray}
\frac{d\text{[ASC\textsubscript{f}]}}{dt}&=&-k_{3}\ F([\text{NLRP3\textsubscript{o}}])\ \text{[NLRP3\textsubscript{o}]}\ \text{[ASC\textsubscript{f}]}, \label{eqA}\\
\frac{d\text{[ASC\textsubscript{b}]}}{dt}&=&k_{3}\ F([\text{NLRP3\textsubscript{o}}])\ \text{[NLRP3\textsubscript{o}]}\ \text{[ASC\textsubscript{f}]},\label{eqNA}
\end{eqnarray}
where the function $F([\text{NLRP3\textsubscript{o}}])$ is included in order to model the absence or presence of the NLRP3 inflammasome base. 
In order to ensure that binding of the ASC occurs only once the inflammasome base has formed in our continuous model, we approximate this two-state system as a continuous sigmoid function \cite{step_sigm_iliev} such that,
\begin{equation}
 F(\text{[NLRP3\textsubscript{o}])}=\frac{1}{ 1+ \Big( \frac{ \text{[NLRP3\textsubscript{o}]} + a}{b} \Big) ^{-c} }.
\label{eq:F_cont}
\end{equation}
Here, values $a=1$ (a.u.) and $b=2$ (a.u.) can be used to obtain $F([\text{NLRP3\textsubscript{o}}]) \approx 0$ for $[\text{NLRP3\textsubscript{o}}] < 1$ and $F(\text[\text{NLRP3\textsubscript{o}}]) \approx 1$ otherwise. The value of $c$ is set to be large ({\em e.g.,} 1000) in order to create a steep curve in the sigmoid. These values result in the binding of ASC once $[\text{NLRP3\textsubscript{o}}]$ reaches a threshold value $n\approx 1$ a.u., that is when oligomerised NLRP3 reaches this level the inflammasome is formed.
{Note that the total amount of ASC is conserved in our model,} 
\begin{equation}
\text{[ASC\textsubscript{f}]}+\text{[ASC\textsubscript{b}]}=\text{constant}.
\label{conserve2}
\end{equation}
Free ASC levels are adequate for pyroptosis at homeostasis~\cite{Kozloski2020,Bertinaria2019}. Therefore, in an effort to minimise model complexity, we do not include the synthesis or decay of ASC in the model.

\subsection{Caspase-1 dynamics}
\label{sec:model_caspase1}
Pro-caspase-1 can bind to inflammasome-bound ASC upon availability, and subsequently dimerise into its activated, mature form, caspase-1. In our model, concentrations of pro-caspase-1 and caspase-1 are denoted by [pro-C1] and [C1], respectively. Caspase-1 activation occurs at a rate $k_{4}$ and is here assumed to be irreversible so that the rate of change of [pro-C1] and [C1] can be described by,
\begin{eqnarray}
\frac{d\text{[pro-C1]}}{dt}&=&-k_{4}\ \text{[ASC\textsubscript{b}]}\ \text{[pro-C1]},\label{eqPc}\\
\frac{d\text{[C1]}}{dt}&=&k_{4}\ \text{[ASC\textsubscript{b}]}\ \text{[pro-C1]},\label{eqC}
\end{eqnarray}
so that,
\begin{equation}
\text{[pro-C1]}+\text{[C1]}=\text{constant}.
\label{conserve3}
\end{equation}
Pro-caspase-1 is present in adequate levels in the cell prior to cellular responses to DAMPs or PAMPs~\cite{Kozloski2020,Bertinaria2019}. Therefore, to minimise model complexity, neither the synthesis nor the decay of caspase-1 is included in the model. 

\subsection{Gasdermin dynamics}
\label{sec:model_gsdmd}
GSDMD-N, which is formed as a result of the caspase-1 cleavage of GSDMD, produces pores on the cell membrane that are central to the pyroptosis process. In the model, we consider the concentrations of both GSDMD and GSDMD-N, denoted as [GSDMD] and [GSDMD-N], respectively. We describe the caspase-1 facilitated cleavage using a Hill function, where $\gamma_{\text{C1}}$ is the Hill coefficient, and C1\textsubscript{50} is the half-max [C1] value. We additionally assume a specific rate for [GSDMD] cleavage, $\alpha_{2}$. Therefore, we describe the rate of change of [GSDMD] and [GSDMD-N] in the model, through the equations,
\begin{eqnarray}
\frac{d\text{[GSDMD]}}{dt}&=&-\alpha_{2}\ \frac{\text{[C1]}^{\gamma_{C1}}}{\text{C1}_{50}^{\gamma_{C1}}+\text{[C1]}^{\gamma_{C1}}}\ \text{[GSDMD]},\label{eqPg}\\
\frac{d\text{[GSDMD-N]}}{dt}&=&\alpha_{2}\ \frac{\text{[C1]}^{\gamma_{C1}}}{\text{C1}_{50}^{\gamma_{C1}}+\text{[C1]}^{\gamma_{C1}}}\ \text{[GSDMD]},\label{eqG}
\end{eqnarray}
so that the following conservation law holds,
\begin{equation}
\text{[GSDMD]}+\text{[GSDMD-N]}=\text{constant}.
\label{conserve4}
\end{equation}
Adequate levels of GSDMD required for pyroptosis are available in the cell at homeostasis~\cite{broz2019gasdermins}, therefore GSDMD synthesis and decay are omitted in the model. 

\subsection{Cytokine dynamics}
\label{sec:model_il1b8}
Translocation of NF-$\kappa$B, from the cytoplasm to the nucleus, induces the transcription and synthesis of pro-IL-1$\beta$, but does not up-regulate pro-IL-18 activity from homeostatic levels. When active caspase-1 is available, the pro-forms of the interleukins can be cleaved into their activated forms. Subsequently, once membrane pores have been formed in response to GSDMD-N activity, cytoplasmic interleukins can be secreted into the extracellular region. We here consider the concentrations of IL-1$\beta$/18 in pro-, cytoplasmic and extracellular form, respectively denoted [pro-IL-1$\beta$/18], [IL-1$\beta$\textsubscript{c}/18\textsubscript{c}] and [IL-1$\beta$\textsubscript{e}/18\textsubscript{e}]. 
In the model, pro-IL-1$\beta$ and pro-IL-18 are cleaved by caspase-1 at rates $\alpha_{4}$ and $\alpha_{5}$, respectively. Once cleaved, cellular IL-1$\beta$\textsubscript{c} and IL-18\textsubscript{c} are secreted through the GSDMD-N derived pores at the rates $k_{5}$ and $k_{6}$, respectively. 

\paragraph{IL-1$\beta$:}
The transcription and synthesis of pro-IL-1$\beta$ is promoted by [NF-$\kappa$B\textsubscript{n}], in a similar way to NLRP3, therefore we describe this process using a Hill function with constant coefficient $\alpha_{3}$, Hill coefficient $\gamma_{\text{NF}}$ and the half-max concentration of NF-$\kappa$B, NF\textsubscript{50}. We additionally consider the decay of IL-1$\beta$ at the rate $\delta_{2}$. We incorporate the above mechanisms to describe the rate of change of [pro-IL-1$\beta$], [IL-1$\beta$\textsubscript{c}] and [IL-1$\beta$\textsubscript{e}] over time as,
\begin{eqnarray}
\frac{d\text{[pro-IL-1}\beta]}{dt}&=&\alpha_{3}\ \frac{[\overline{\text{NF-$\kappa$B\textsubscript{n}}}]^{\gamma_{NF}}}{\text{NF}_{50}^{\gamma_{NF}}+[\overline{\text{NF-$\kappa$B\textsubscript{n}}}]^{\gamma_{NF}}}-\alpha_{4}\ \frac{\text{[C1]}^{\gamma_{C1}}}{\text{C1}_{50}^{\gamma_{C1}}+\text{[C1]}^{\gamma_{C1}}}\ \text{[pro-IL-1$\beta$]}\nonumber\\
&&
-\delta_{2}\ \text{[pro-IL-1$\beta$]},\label{eqP1b}\\
\frac{d\text{[IL-1$\beta$\textsubscript{c}]}}{dt}&=& \alpha_{4}\ \frac{\text{[C1]}^{\gamma_{C1}}}{\text{C1}_{50}^{\gamma_{C1}}+\text{[C1]}^{\gamma_{C1}}}\ \text{[pro-IL-1$\beta$]}-k_{5}\ G\ \text{[IL-1$\beta$\textsubscript{c}]}\nonumber\\
&&-\delta_{2}\ \text{[IL-1$\beta$\textsubscript{c}]},\label{eq1bc}\\
\frac{d\text{[IL-1$\beta$\textsubscript{e}]}}{dt}&=&k_{5}\ G \text{ [IL-1$\beta$\textsubscript{c}]},\label{eq1be}
\end{eqnarray}
where $[\overline{\text{NF-$\kappa$B\textsubscript{n}}}]$ is defined in Equation~\eqref{eq:nfkb_shift_for_transcription} and $G$ is a non-dimensionalised value that allows for [GSDMD-N]-dependent cytokine outflux, and water influx, such that,
\begin{equation}
 G=G(\text{[GSDMD-N]}):= \frac{\text{[GSDMD-N]}}{\text{[GSDMD]+[GSDMD-N]}}.
 \label{eqGFLUX}
\end{equation}
\paragraph{IL-18:}
We describe the rate of change of [pro-IL-18], [IL-18\textsubscript{c}] and [IL-18\textsubscript{e}] as,
\begin{eqnarray}
\frac{d\text{[pro-IL-18]}}{dt}&=&-\alpha_{5}\ \frac{\text{[C1]}^{\gamma_{C1}}}{\text{C1}_{50}^{\gamma_{C1}}+\text{[C1]}^{\gamma_{C1}}}\ \text{[pro-IL-18]},\label{eqP18}\\
\frac{d\text{[IL-18\textsubscript{c}]}}{dt}&=& \alpha_{5}\ \frac{\text{[C1]}^{\gamma_{C1}}}{\text{C1}_{50}^{\gamma_{C1}}+\text{[C1]}^{\gamma_{C1}}}\ \text{[pro-IL-18]}-k_{6}\ G\ \text{[IL-18\textsubscript{c}]},\label{eq18c}\\
\frac{d\text{[IL-18\textsubscript{e}]}}{dt}&=&k_{6}\ G\ \text{[IL-18\textsubscript{c}]},\label{eq18e}
\end{eqnarray}
so that,
\begin{equation}
\text{[pro-IL-18]}+\text{[IL-18\textsubscript{c}]}+\text{[IL-18\textsubscript{e}]}=\text{constant}.
\label{conserve5}
\end{equation}
The levels of pro-IL-18 in the cell are kept at a homeostatic level~\cite{Kozloski2020,Bertinaria2019} and thus the synthesis and decay of IL-18 are not include in the model. 

\subsection{Cell volume}
\label{sec:model_volume}
The GSDMD-N derived membrane pores that allow for the outflux of mature interleukins, also allow for the influx of extracellular material ({\em e.g.}, water). This influx causes the cell to swell until its membrane eventually ruptures. Single-cell analysis has revealed that before the ultimate membrane rupture occurs, the cell volume increases gradually~\cite{deVasconcelos2019}. Thus, in the model, we consider the volume of the cell, $V$, to increase at a rate $k_{7}G$ once pores are formed on the cell membrane. We describe the rate of change of the cell volume by the equation,
\begin{eqnarray}
\frac{dV}{dt}&=&k_{7}\ G\ V.\label{eqV}
\end{eqnarray}
Once the cell volume reaches a critical volume $V_{c}$, the cell ruptures and all cell processes cease.

\subsection{Modified drug targeting model}
\label{sect:drug}
We consider a generic anti-inflammatory drug (Drug) which binds to active NLRP3 and thus inhibits the inflammasome formation by preventing NLRP3\textsubscript{a} oligomerisation. To incorporate this in our model, we consider the drug binding to NLRP3\textsubscript{a} to form the complex Drug $\cdot$ NLRP3\textsubscript{a}. The forward and reverse rate constants for this reaction are denoted by $k_{+\text{D}}$ and $k_{-\text{D}}$, respectively. 
When including these drug mechanisms in our model, Equation~\eqref{eqNa} describing the dynamics of NLRP3\textsubscript{a} must be modified to include two additional terms,
\begin{equation}
 \frac{d\text{[NLRP3\textsubscript{a}]}}{dt}=\ldots-k_{+\text{D}}\ \text{[Drug]}\ \text{[NLRP3\textsubscript{a}]} + k_{-\text{D}}\ \text{[Drug $\cdot$ NLRP3\textsubscript{a}}]. \label{eqNa_drug}
\end{equation}
The rate of change of [Drug] and [Drug $\cdot$ NLRP3\textsubscript{a}] can now be included in the model as,
\begin{eqnarray}
\frac{d\text{[Drug]}}{dt}&=& - k_{+\text{D}}\ \text{[Drug]}\ \text{[NLRP3\textsubscript{a}]} + k_{-\text{D}}\ \text{[Drug $\cdot$ NLRP3\textsubscript{a}]}, \label{eqDrug_1b} \\
\frac{d\text{[Drug $\cdot$ NLRP3\textsubscript{a}]}}{dt}&=& k_{+\text{D}}\ \text{[Drug]}\ \text{[NLRP3\textsubscript{a}]} - k_{-\text{D}}\ \text{[Drug $\cdot$ NLRP3\textsubscript{a}]}, \label{eqDrug_1}
\end{eqnarray}
where the following conservation law holds,
\begin{equation}
\label{eqconDrug}
 \text{[Drug]+[Drug $\cdot$ NLRP3\textsubscript{a}] = constant.}
\end{equation}

\subsection{Implementation of the pathway model}
\label{sect:implement}
In order to reduce the pyroptosis model, as described in Sections~\ref{sec:model_nfkb}-\ref{sect:drug}, we impose the conservation laws~\eqref{conserve2}, \eqref{conserve3}, \eqref{conserve4}, \eqref{conserve5} and \eqref{eqconDrug}. We further set all total concentration constants to be 1 arbitrary unit of concentration (1 a.u.). Below, we provide the reduced system of equations, 
\begin{eqnarray}
\text{[NF-$\kappa$B\textsubscript{n}]}({t}) &=& \text{[NF-$\kappa$B\textsubscript{n}]}(0) + S_1 \cdot h\ e^{- \frac{\log^2 (t/\tau)}{s} },\nonumber\\
\frac{d\text{[NLRP3\textsubscript{i}]}}{dt}&=&\alpha_1 \text{Hill}_{\text{NF}}-S_2 k_{1}\text{[NLRP3\textsubscript{i}]}-\delta_1 \text{[NLRP3\textsubscript{i}]}, \nonumber\\
\frac{d\text{[NLRP3\textsubscript{a}]}}{dt}&=&S_2 k_{1}\text{[NLRP3\textsubscript{i}]}-k_{2} \text{[NLRP3\textsubscript{a}]}^{2}-\delta_1 \text{[NLRP3\textsubscript{a}]}\nonumber\\
&&-k_{+\text{D}}\ \text{[Drug]}\ \text{[NLRP3\textsubscript{a}]} + k_{-\text{D}}\ \text{[Drug $\cdot$ NLRP3\textsubscript{a}}], \nonumber\\
\frac{d\text{[NLRP3\textsubscript{o}]}}{dt}&=&k_{2} \text{[NLRP3\textsubscript{a}]}^{2}, \nonumber\\
\frac{d\text{[ASC\textsubscript{b}]}}{dt}&=&k_{3}\ F_{\text{N}_{\text{o}}}\ \text{[NLRP3\textsubscript{o}]}\ (1-\text{[ASC\textsubscript{b}]}), \nonumber\\
\frac{d\text{[C1]}}{dt}&=&k_{4}\ \text{[ASC\textsubscript{b}]}\ (1-\text{[C1]}), \nonumber\\
\frac{d\text{[GSDMD-N]}}{dt}&=&\alpha_{2}\ \text{Hill}_{\text{C1}}\ (1-\text{[GSDMD-N]}),\nonumber\\
\frac{d\text{[pro-IL-1}\beta]}{dt}&=&\alpha_{3}\ \text{Hill}_{\text{NF}} -\alpha_{4}\ \text{Hill}_{\text{C1}} \text{[pro-IL-1$\beta$]}-\delta_{2}\ \text{[pro-IL-1$\beta$]},\label{full_system}\\
\frac{d\text{[IL-1$\beta$\textsubscript{c}]}}{dt}&=& \alpha_{4}\ \text{Hill}_{\text{C1}} \text{[pro-IL-1$\beta$]}-k_{5} G \text{[IL-1$\beta$\textsubscript{c}]}-\delta_{2}\ \text{[IL-1$\beta$\textsubscript{c}]},\nonumber\\
\frac{d\text{[IL-1$\beta$\textsubscript{e}]}}{dt}&=&k_{5}\ G \text{[IL-1$\beta$\textsubscript{c}]}, \nonumber\\
\frac{d\text{[IL-18\textsubscript{c}]}}{dt}&=& \alpha_{5} \text{Hill}_{\text{C1}} (1-\text{[IL-18\textsubscript{c}]}-\text{[IL-18\textsubscript{e}]})-k_{6}G \text{[IL-18\textsubscript{c}]},\nonumber\\
\frac{d\text{[IL-18\textsubscript{e}]}}{dt}&=&k_{6} G \text{[IL-18\textsubscript{c}]}, \nonumber\\
\frac{d\text{V}}{dt}&=&k_{7}G \text{V},\nonumber\\
\frac{d\text{[Drug]}}{dt}&=& - k_{+\text{D}}\ \text{[Drug]}\ \text{[NLRP3\textsubscript{a}]} + k_{-\text{D}}\ \text{[Drug $\cdot$ NLRP3\textsubscript{a}]}, \nonumber \\
\frac{d\text{[Drug $\cdot$ NLRP3\textsubscript{a}]}}{dt}&=& k_{+\text{D}}\ \text{[Drug]}\ \text{[NLRP3\textsubscript{a}]} - k_{-\text{D}}\ \text{[Drug $\cdot$ NLRP3\textsubscript{a}]},\nonumber
\end{eqnarray}
with the following definitions,
\begin{eqnarray}
\text{Hill}_{\text{NF}}&=&\frac{(\text{[NF-$\kappa$B\textsubscript{n}]}(t)-\text{[NF-$\kappa$B\textsubscript{n}]}(0))^{\gamma_{\text{NF}}}}{\text{NF}_{50}^{\gamma_{\text{NF}}}+(\text{[NF-$\kappa$B\textsubscript{n}]}(t)-\text{[NF-$\kappa$B\textsubscript{n}]}(0))^{\gamma_{\text{NF}}}},\nonumber\\
F_{\text{N}_{\text{o}}}&=&\frac{1}{1+\left( \frac{\text{[NLRP3\textsubscript{o}]}-a}{b}\right)^{-c}},\label{full_functions}\\
\text{Hill}_{\text{C1}}&=&\frac{\text{[C1]}^{\gamma_{C1}}}{\text{C1}_{50}^{\gamma_{C1}}+\text{[C1]}^{\gamma_{C1}}}.\nonumber
\end{eqnarray}

The reduced model is implemented in {{\sc MATLAB}~\cite{MATLAB:2019b}}, where the system of ODEs,~\eqref{full_system} with~\eqref{full_functions}, is solved numerically using the built-in ODE-solver \texttt {ode15s}. Model outputs are measured in terms of arbitrary units of concentration or volume over time. 
The model parameters used in the simulations, and the justifications for these values, are provided in Table~\ref{tab:par} and Appendix~\ref{app_parameters}, respectively. 

The initial conditions pertaining to NF-$\kappa$B are motivated by the works of Bagaev {\em et al.}~\cite{bagaev2019elevated}, reporting initial nuclear NF-$\kappa$B concentrations to be between 10\% and 30\% (of the total amount of cellular NF-$\kappa$B), depending on cell line. Therefore we choose,
$$
[\text{NF-$\kappa$B}_{\text{n}}](0)=0.25 \text{ a.u.}, 
$$
where the total amount of cellular NF-$\kappa$B is set to be  1 a.u.. 
Similarly, we set,
$$
[\text{ASC\textsubscript{f}}](0)=1 \text{ a.u.},\ [\text{pro-C1}](0)= 1 \text{ a.u.},
$$
$$
\ \text{[GSDMD]}(0)=1 \text{ a.u.},\ \text{[pro-IL-18]}(0)=1 \text{ a.u.},\ V(0)=1 \text{ a.u.},
$$
while all other initial concentrations are set to be 0 a.u.. For the results shown in Figure~\ref{fig:basecase}, we consider the model without drug intervention, whereas, in the results shown in Figure~\ref{fig:completedrug}, we consider the initial concentration of the drug to be a value between 0 and 1 a.u.. Note, as we consider an {\em in vitro} situation, we consider the drug to be present from the start of the simulations and not decay within the simulation time. 

A more detailed description of the numerical set-up can be found in Appendix~\ref{app:code_and_numerics}, which also includes instructions on how to access and run the code.
Simulation results are provided and discussed in the next section.

\subsection{Results and discussion of the pathway model}
\label{sect:results}
\begin{figure}
\begin{center}
 \includegraphics[width=\textwidth]{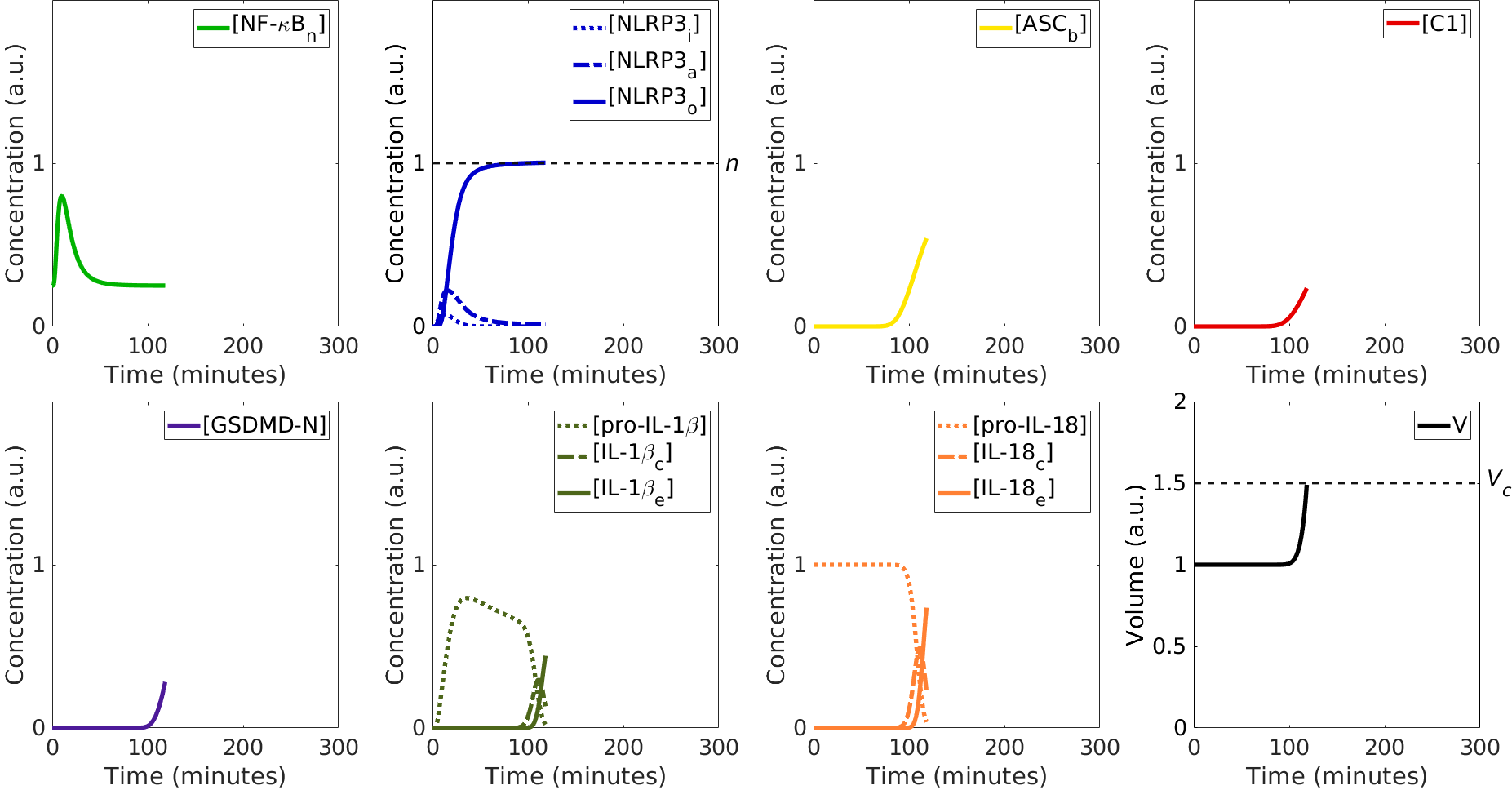}
 \caption{Numerical simulations results of the system of ODEs~\eqref{full_system} with functions~\eqref{full_functions} and [Drug](0)=0 a.u., that is, the model of pyroptosis in the absence of the anti-inflammatory drug. We display the concentration of each model component in arbitrary units (a.u.) over time, as well as cell volume dynamics. Note that the time progression stops once the critical volume $V_c$ is reached and the ultimate cell membrane rupture occurs.
 }
 \label{fig:basecase}
 \end{center}
\end{figure}

We first consider the model without any drug, that is, we simulate the system of ODEs~\eqref{full_system} with functions~\eqref{full_functions} and [Drug](0)=0 a.u.. The simulation results are provided in Figure~\ref{fig:basecase}, in which the molecule concentrations (in arbitrary units) and cell volume are plotted over time. 
The results show that NF-$\kappa$B is rapidly translocated to the nucleus upon DAMP/PAMP-induced TLR signalling. 
After [NF-$\kappa$B\textsubscript{n}] levels have peaked, the transcription, and by extension the synthesis, of NLRP3\textsubscript{i} and pro-IL-1$\beta$ are reduced as NF-$\kappa$B leaves the nucleus. 
Synthesised NLRP3\textsubscript{i} activates to become NLRP3\textsubscript{a}, which in turn oligomerises and binds together to form the inflammasome base, here expressed in terms of the concentration of NLRP3\textsubscript{o}. These dynamics are captured in the subplot displaying the time evolution of [NLRP3\textsubscript{i}], [NLRP3\textsubscript{a}] and [NLRP3\textsubscript{o}].
Note that once the inflammasome base is formed, the oligomerised NLRP3 concentration plateaus and the binding of ASC to the inflammasome increases. 
This can be observed in the results, as [ASC\textsubscript{b}] levels increase when [NLRP3\textsubscript{o}] reaches the value $n$, where here $n=1$ a.u.. 
Furthermore, as [ASC\textsubscript{b}] levels increase, the dimerisation and activation of caspase-1 is facilitated so that [C1] levels increase. 
In turn, when [C1] levels increase, the cleavage of GSDMD, pro-IL-1$\beta$ and pro-IL-18 occurs. This is apparent in the subplots, as [GSDMD-N], [IL-1$\beta$\textsubscript{c}] and [IL-18\textsubscript{c}] levels start increasing once [C1] levels surpass zero a.u.. 
The outflux of mature interleukins and the influx of extracellular water to the cell are mediated by GSDMD-N derived pores.
Thus, once GSDMD-N has been cleaved, so that [GSDMD-N]$>$0 a.u., the cytoplasmic interleukin concentrations [IL-1$\beta$\textsubscript{c}] and [IL-18\textsubscript{c}] respectively decrease in favour of the extracellular concentrations [IL-1$\beta$\textsubscript{e}] and [IL-18\textsubscript{e}]. 
Increasing [GSDMD-N] levels also enable the cellular volume to increase, which eventually results in membrane rupture once the volume reaches the maximal capacity $V_c$, where here $V_c=1.5$ a.u.. 
Note that when membrane rupture occurs, all intracellular mechanisms are ceased and thus the time progression in the subplots stops. 

\begin{figure}
\begin{center}
 \includegraphics[width=\textwidth]{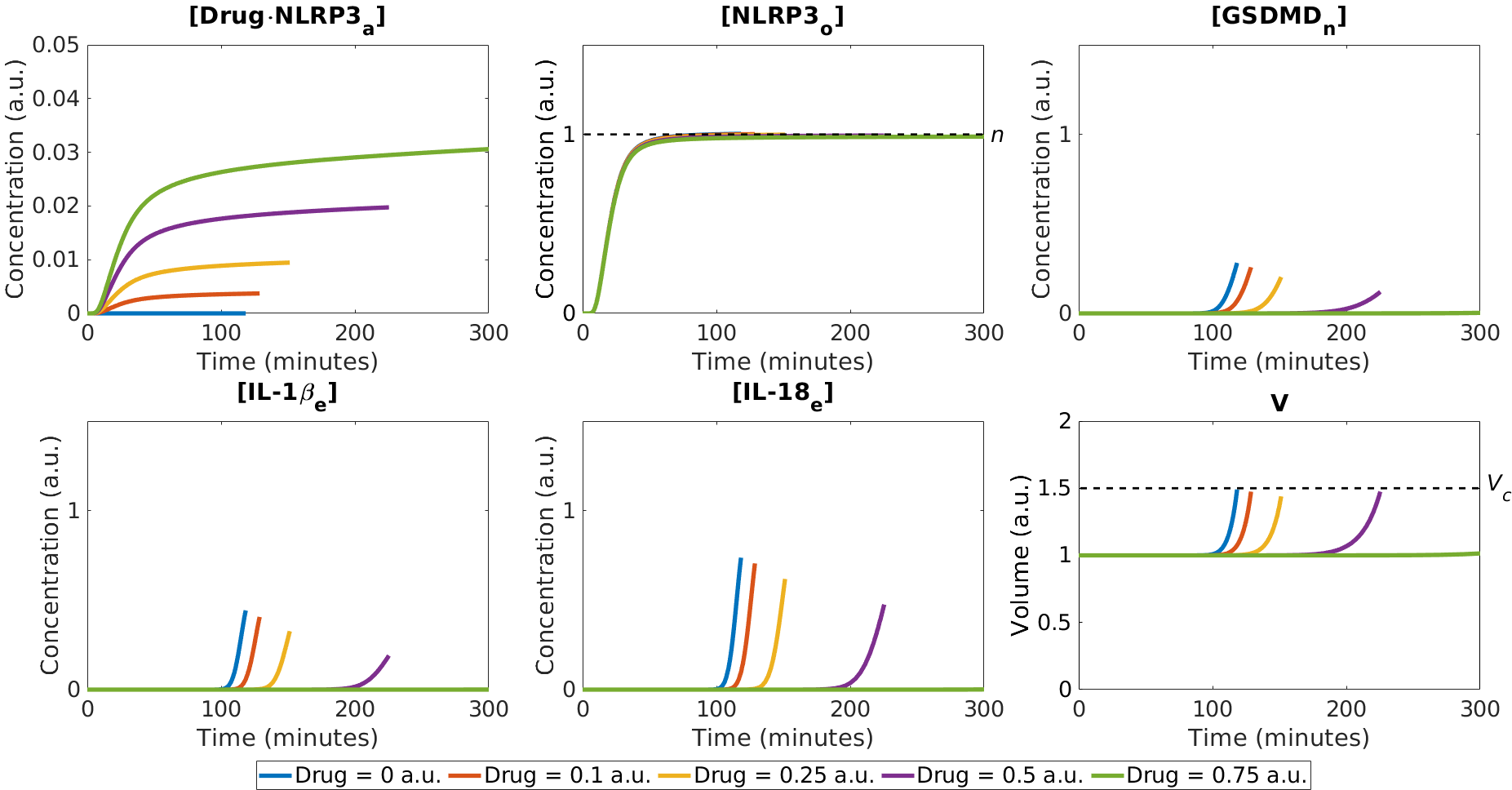}
 \caption{Numerical simulations results of the system of ODEs~\eqref{full_system} with functions~\eqref{full_functions} with non-zero drug initial concentrations. We display the concentration of each model component in arbitrary units (a.u.) over time, as well as cell volume dynamics for several dosages of the drug. The dynamics of the key model components are plotted, where the line colour corresponds to applied drug dosage, as described in the legend. Note that the time progression stops if the critical volume ({\em i.e.,} $V_c=1.5$ a.u.) is reached, as the cell membrane ruptures.}
 \label{fig:completedrug}
 \end{center}
\end{figure}

We next consider numerical simulations of the model including the effects of an anti-inflammatory drug, that is system of ODEs~\eqref{full_system} with functions~\eqref{full_functions} and [Drug](0)$>$0 a.u.. The results of the numerical simulations are displayed in Figure~\ref{fig:completedrug}, where we compare different initial concentrations (dosages) of the drug. 
We display the time evolution of each model component in a separate subplot, where the graph colour corresponds to drug dosage. 
These results show that increasing levels of the drug push pyroptotic events further forward in time. Note that the results for the maximal tested drug dosage results in many of the processes driving pyroptosis not occurring in the time frame examined. 
The drug specifically inhibits the NLRP3 oligomerisation process. 
Therefore, when it is added to the system, we observe that as the drug concentration [Drug] increases, so does the amount of time it takes for [NLRP3\textsubscript{o}] to reach the threshold value $n$. 
This means that the inflammasome base formation, and by extension the downstream processes resulting in inflammatory cytokine secretion and the ultimate membrane rupture, are delayed.

\FloatBarrier

\subsection{Sensitivity analyses of the pathway model}
\label{sect:sens}
The mathematical model of the pyroptosis pathway includes the subset of model parameters $\Theta$ =\{$\theta_1$, $\theta_2$,...,$\theta_p$\}, which are listed in Table~\ref{tab:par} in Appendix~\ref{app_parameters}. 
As is described in Appendix~\ref{app_parameters}, some model parameter values are directly obtained from published data~\cite{han2015lipopolysaccharide,moors2000proteasome}, some are estimated from immunoblot data~\cite{Huang2018}, whilst other values are collectively approximated through data fitting to time-course data that are available in the literature~\cite{bagaev2019elevated,deVasconcelos2019,martin2016inflammasome}. We note that such a data fitting approach may lead to large parameter uncertainties, and thus we conduct sensitivity analyses in order to investigate how sensitive the model outputs are to changes in the values of the model parameters $\Theta$, hereby referred to as the model inputs. 
Such analyses can help us understand how changes in the model input can affect the model output~\cite{qian_mahdi_2020,uncert_sens_review_Paper}. Therefore, this can allow us to identify which input parameters $\theta_j$ are the most influential, providing information on the importance of experimental retrieval of accurate parameters and possible model reductions.
In this study, we investigate the system of ODEs~\eqref{full_system} describing the pyroptosis pathway, using three different sensitivity analysis techniques, which are robustness analysis (Section~\ref{sec:ss_robust}), Latin hypercube analysis (Section~\ref{sec:ss_lhca}) and a derivative-based method (Section~\ref{sec:ss_dirmeth}). The molecule concentrations over time are the model outputs $y_i(t)$, where $i=1,...,q$, as labelled in Table~\ref{tab:var}.

\subsubsection{Robustness analyses}
\label{sec:ss_robust}
\begin{figure}
\begin{center}
 \includegraphics[width=\textwidth]{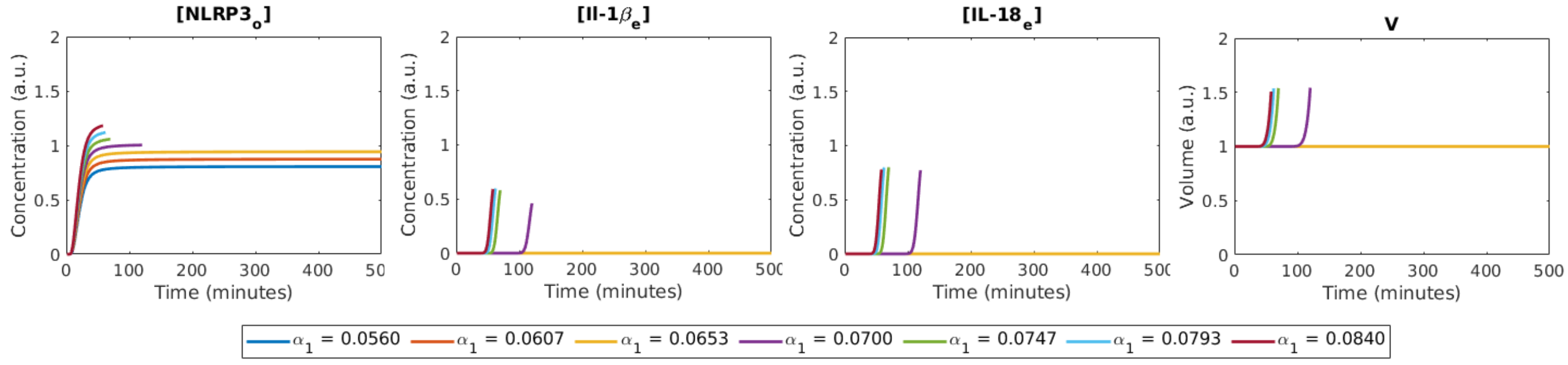}
 \caption{Robustness analyses. The plots show numerical simulations results of the ODE system~\eqref{full_system}, where the input parameter $\alpha_1$ (in a.u. min$^{-1}$) is perturbed, whilst other model parameters are fixed at their estimated values. Model components are measured in terms of arbitrary concentration units (a.u.) or volume. 
 Note that the time progression stops once the critical volume $V_c$=1.5 a.u. is reached and the ultimate cell membrane rupture occurs.
 } 
 \label{fig:sensnorupture}
 \end{center}
\end{figure}

Robustness analysis is a local sensitivity analysis technique in which one input parameter $\theta_j$ is varied, whilst the other $p-1$ parameters are kept fixed at their estimated values~\cite{qian_mahdi_2020,uncert_sens_review_Paper}.
We compare the output molecule concentrations $y_i(t)$ over time for 7 different perturbed values of each parameter $\theta_j$, where the perturbed values range between $\pm20\%$ of the estimated value. 
In Figure~\ref{fig:sensnorupture}, we include robustness analysis results for one input parameter, specifically $\alpha_1$, which denotes the rate coefficient for the transcription of inactive NLRP3. This figure shows that increasing the parameter value of $\alpha_1$ speeds up the inflammasome base formation and, thereby, the pyroptosis process. For investigated values of $\alpha_1$ below the threshold value of $\alpha_1=0.07$ (a.u.) min$^{-1}$, the pyroptosis process is delayed to an extent that NLRP3 formation does not occur within the simulated 500 minutes. 
For the estimated values of the parameter set $\Theta$, we consider such perturbations of $\alpha_1$ to be unfeasible, since we are here aiming to simulate a scenario in which the NLRP3 inflammasome base is formed at around 77 minutes, and the ultimate membrane rupture occurs at 120 minutes, as is described in Appendix~\ref{app_parameters}. Robustness analysis results for all model inputs $\theta_j,j=1,...,21$ and model outputs $y_i(t)$, $i=1,...,15$ are available in the Supplementary Material (Supplementary Material, S2). These analyses provides a visual overview of how increasing or decreasing each model parameter affects the output. 

\subsubsection{Global sensitivity analysis}
\label{sec:ss_lhca}
\begin{figure}
 \centering
 \includegraphics[width=\textwidth]{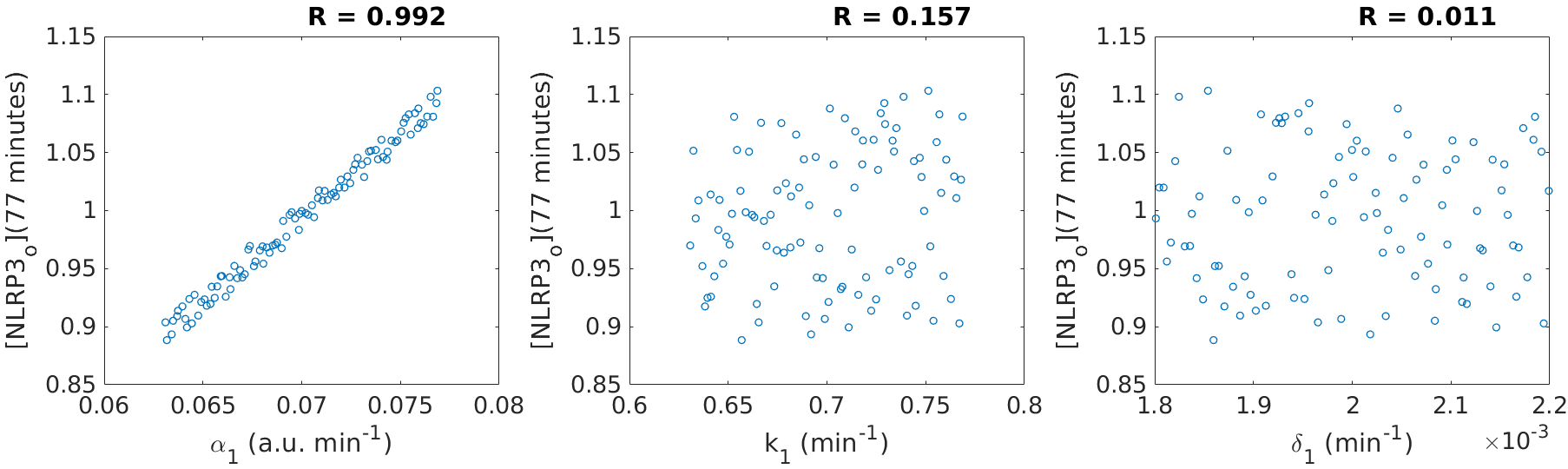}
 \caption{Global sensitivity analysis is performed using Latin hypercube sampling and analysis. The output, {\em{i.e.,}} the NLRP3\textsubscript{o} concentration (in arbitrary units of concentration) at 77 minutes, is plotted over three input parameters. Each input parameter range spans $\pm10\%$ of the estimated parameter value. 
 The Pearson correlation coefficient (R), measuring the linear association between each input-output pair, is provided at the top right corner of each corresponding subplot. }
 \label{fig:Global_sens_LHC} 
\end{figure}

We perform a global sensitivity analysis, in which all parameters $\theta_j$ downstream of NF-$\kappa$B dynamics are simultaneously perturbed from their estimated values, using Latin hypercube sampling and analysis~\cite{qian_mahdi_2020,uncert_sens_review_Paper}.
Results from the Latin hypercube analysis are provided in the form of scatter plots, where the model outputs are plotted over the investigated parameter ranges ($\pm 10\%$ of the estimated parameter values) for each input parameter. 
Here, the regarded output is the NLRP3\textsubscript{o} concentration at 77 minutes (the time point where the inflammasome base is formed {\em i.e.,} when [NLRP3\textsubscript{o}] reaches the value $n=1$ a.u. when using the estimated parameter values). 
The scatter plots in Figure~\ref{fig:Global_sens_LHC} provide a visual means to study the relationship between three model inputs ($\alpha_1, k_1, \delta_1$) and the regarded output, for small perturbations of the input parameters. 
Scatter plots for the other input parameters are available in the Supplementary Material (Supplementary Material, S3). 
In order to quantify linear input-output associations, Figure~\ref{fig:Global_sens_LHC} also includes Pearson correlation coefficients (R) for each input-output pair, evaluated over the regarded input parameter ranges~\cite{uncert_sens_review_Paper}. 
A strong positive linear input-output correlation (R=0.992) is found for input parameter $\alpha_1$, implying that, given the estimated parameter values of $\Theta$, our model results are especially sensitive to small perturbations of $\alpha_1$ which influences the transcription rate of inactive NLRP3. The linear input-output correlation is weakly positive for input parameter $k_1$ (R=0.157), and negligible for $\delta_1$ (R=0.011), within the investigated parameter ranges. 

\subsubsection{Derivative-based sensitivity analysis}
\label{sec:ss_dirmeth}
\begin{figure}
 \centering
 \includegraphics[width=\textwidth]{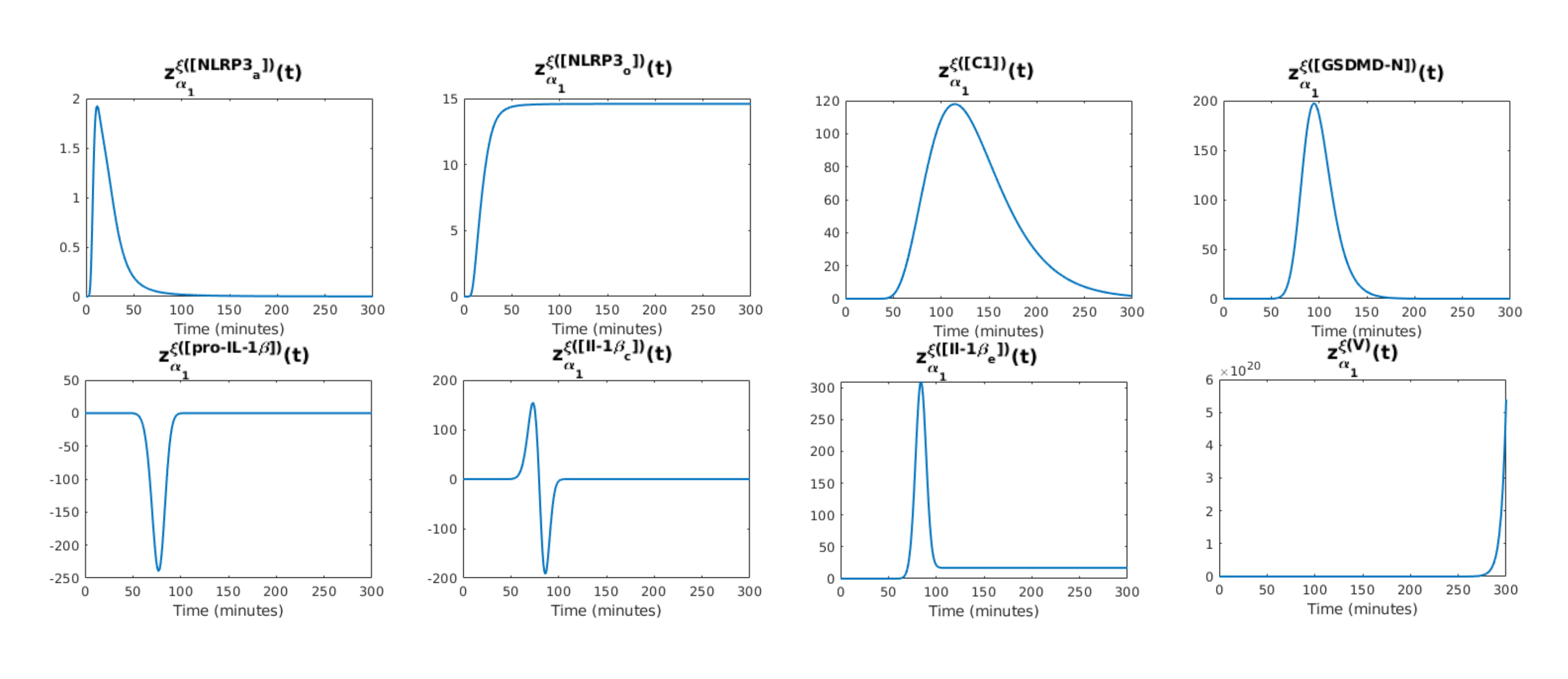}
 \caption{A derivative-based sensitivity analysis shows that various output results are sensitive to perturbations of the input parameter $\alpha_1$, for the estimated parameter values of $\Theta$. Here $z^i_j$ denotes the sensitivity of the output concentration computed in equation $i$, with respect to the input parameter $\theta_j$, where $\xi$(compound concentration) maps each compound to its governing equation index $i$.}
 \label{fig:ss_dm}
\end{figure}

We perform a derivative-based sensitivity analysis, in which the equations describing the time evolution of the concentrations $y_i(t)$, where $i=1,...,q$, are differentiated with respect to each input parameter $\theta_j$ where $j=1,...,p$~\cite{qian_mahdi_2020,dickinson_1976}. 
We thus compute $q \times p$ partial derivatives $z^i_j(t) = \partial y_i(t) / \partial \theta_j$, at each time-point $t$ in which the system of ODEs~\eqref{full_system} is numerically solved. These partial derivatives $z^i_j(t)$ are also referred to as the model sensitivities, and are here numerically computed using the Direct Method, as described by Dickinson and Gelinas~\cite{dickinson_1976}. The values $z^i_j(t)$ provide time-varying measurements of the sensitivity of the outputs $y_i(t)$ with respect to changes in values of the input parameters $\theta_j$. 
Results from the derivative-based sensitivity analysis are presented in Figure~\ref{fig:ss_dm}, in which output sensitivities with respect to $\alpha_1$, {\em i.e,} $z^i_{\alpha_1}$, are plotted over time. 
Results for sensitivities with respect to other model input parameters are available in the Supplementary Material (Supplementary Material, S4). 
Figure~\ref{fig:ss_dm} shows that by increasing $\alpha_1$, the NLRP3\textsubscript{a} levels increase at the start of the simulation (for times under $\sim$ 100 minutes). NLRP3\textsubscript{o} levels also increase with $\alpha_1$, where the magnitude of this increase plateaus when the inflammasome base is formed. 
Thus the outputs [NLRP3\textsubscript{a}] and [NLRP3\textsubscript{o}] are both sensitive to variations of $\alpha_1$. 
As a downstream effect of this, the levels of C1 and GSDMD-N also increase with $\alpha_1$ which, in turn, alter the composition of pro-, cytoplasmic, and external IL-1$\beta$ over time, as well as alter the cell volume $V$. 

\section{Integrating the pyroptosis pathway model with the PhysiCell framework}
\label{sect:abm}
\subsection{Overview}
Our single-cell pyroptosis model, described in Section~\ref{sect:model}, has been integrated within a multiscale tissue simulator that has been developed to investigate within-host dynamics in response to SARS-CoV-2~\cite{wang2020rapid}. This work is part of a multidisciplinary, international coalition led by Paul Macklin (Indiana University, USA). The coalition is made up of several subgroups who are developing individual submodels which are integrated to form a virtual lung epithelium within PhysiCell, an open source physics-based cell simulator~\cite{physicellsimulator}. In the project, epithelial cells are modelled as susceptible to SARS-CoV-2 infection. Other cell types, such as immune cells, are also included in the model but are not susceptible to the virus.

The PhysiCell SARS-CoV-2 framework consists of a hybrid agent-based model that aims to capture; virus transmission in the tissue, intracellular viral processes ({\emph{e.g.}}, binding, endocytosis, replication, exocytosis), infected cell responses ({\emph{e.g.}}, metabolism, chemokine secretion, cell death via apoptosis, cell death via pyroptosis), immune cell dynamics ({\emph{e.g.}}, recruitment, activation, infiltration and predation of infected cells) and tissue damage due to infection and/or host responses. For more details on each model component, we refer the reader to the most recent preprint from the coalition~\cite{wang2020rapid}. In line with a rapid prototyping and open source research approach, the project has been through several versions and will be updated further as each submodel is refined. Here, we will give a brief overview of the hybrid framework and discuss some of the implications of pyroptosis in the PhysiCell tissue simulator, using the version 3 code, with our added pyroptosis model. An initial iteration of our pyroptosis model is integrated into the version 4 tissue simulator code, and the refined model considered here will be implemented into future versions of the code.

\subsection{Simulation outline and implementation}
At the start of each PhysiCell SARS-CoV-2 tissue simulation, virions (virus particles) are non-uniformly distributed across the epithelial cells. Virions diffuse across the tissue and can be endocytosed via ACE-2 receptors, and are thereafter replicated and released by epithelial cells. If the viral RNA in an epithelial cell surpasses a threshold value $R_{\text{RNA}}$, the epithelial cell undergoes pyroptosis with a probability $\Pi \in [0,1]$, and apoptosis with a probability $1-\Pi$. 
The intracellular pyroptosis model governs the cellular pyroptosis-induced secretion of the cytokines IL-18 and IL-1$\beta$, as well as the increase in cytoplasmic volume due to water influx, and the ultimate cell membrane rupture. 

Once secreted, extracellular IL-18 and IL-1$\beta$ are modelled as diffusible fields across the tissue. 
IL-1$\beta$ has been shown to causes a pyroptotic bystander effect, whereby increased IL-1$\beta$ levels can induce pyroptosis in uninfected, or infected, cells~\cite{Bergsbaken2009,lipinska2014applying,schroder2010inflammasomes,Stutz2009}. 
Therefore, in the model, epithelial cells can internalise extracellular IL-1$\beta$, and if the level of internalised IL-1$\beta$ surpasses a threshold value $R_{\text{IL-1}\beta}$, the pyroptosis cascade is initiated in the cell. 
The diffusion, secretion and uptake of virions and cytokines are handled by built-in PhysiCell functionalities. Furthermore, the PhysiCell framework has an inbuilt apoptosis mechanism which results in cell volume shrinkage until the cell is no longer viable. We include further details of this apoptosis model in Appendix~\ref{app:physicell}. The system of ODEs describing the pyroptosis pathway is numerically solved using the backward Euler method. Further details regarding implementation, as well as information on how to access the code, are provided in Appendix~\ref{app:physicell}.

\subsection{Agent-based model results}
We now utilise the PhysiCell SARS-CoV-2 tissue simulator to investigate the effects of pyroptosis on a cellular-level scale. 
We design {\it in silico} experiments to study research questions related to virus-induced cell death. In these \textit{in silico} experiments, we investigate scenarios in which epithelial cells are exposed to virus in the absence of an immune system, thus simulating \textit{in vitro} experiments. We investigate the impact of two key parameters, $R_{\text{IL-1}\beta}$ and $\Pi$, which are described in the above section. 

\begin{figure}
\begin{center}
 \includegraphics[width=\textwidth]{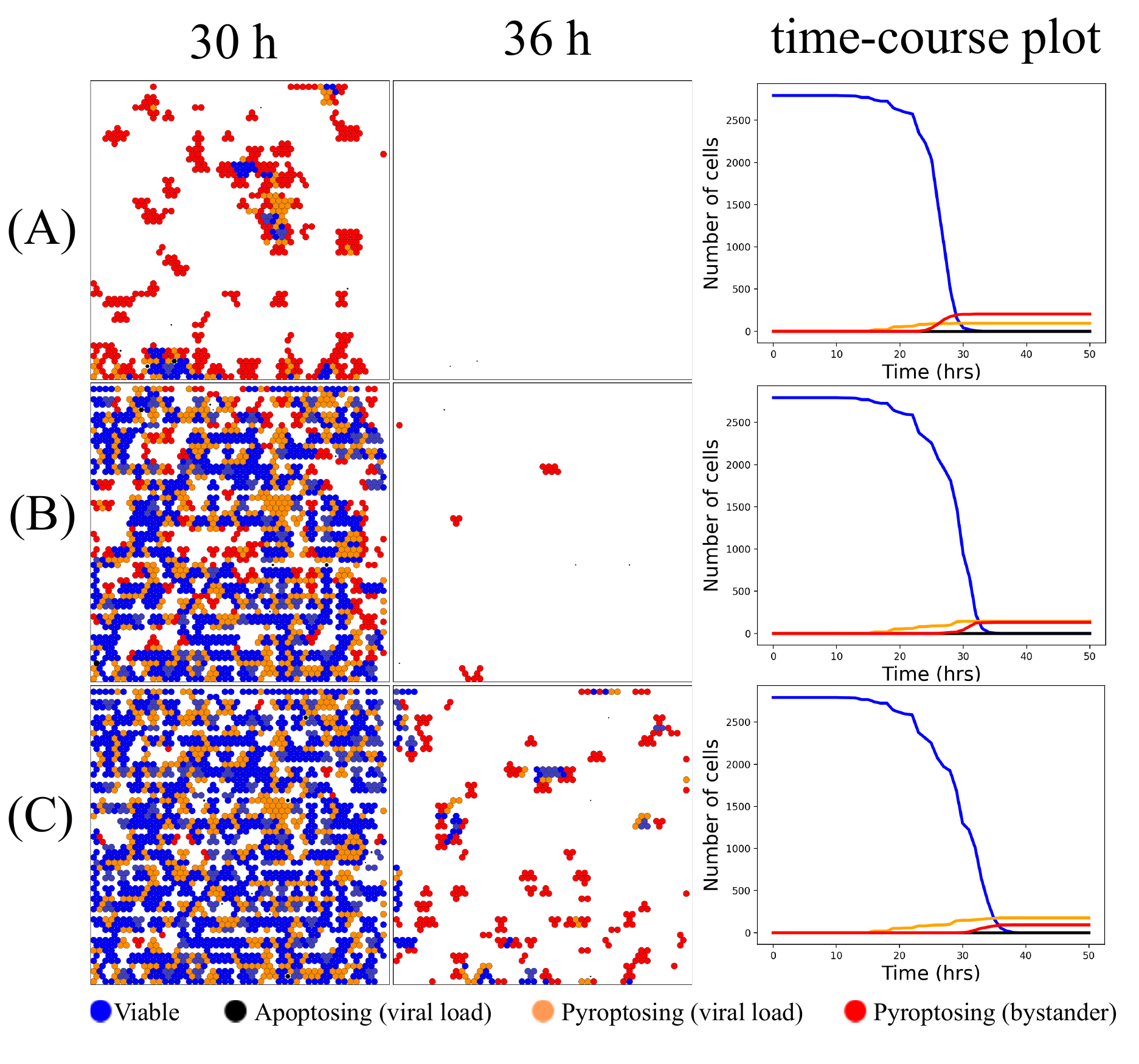}
 \caption{Cell map progressions and averaged time-course data for three situations, investigating the bystander effect of IL-1$\beta$ in the PhysiCell SARS-CoV-2 framework. We test three values for the threshold level of IL-1$\beta$ required to initiate pyroptosis: {\bf (A)} $R_{\text{IL-1}\beta}=0.5\ R_{\text{est}}$, {\bf (B)} $R_{\text{IL-1}\beta}= R_{\text{est}}$ and {\bf (C)} $R_{\text{IL-1}\beta}= 1.5\ R_{\text{est}}$, where $R_{\text{est}}$ is a chosen base-line value. Viable cells (blue), infected cells (darker blue), cells undergoing apoptosis (black), cells undergoing viral load induced pyroptosis (orange) and cells undergoing bystander induced pyroptosis (red) are shown in the cell maps, and the corresponding cell counts at each time-point are displayed in the time-course plots. Note, here $\Pi=1$. 
 }
 \label{fig:PhysiCell_IL1b}
 \end{center}
\end{figure}

\begin{figure}
\begin{center}
 \includegraphics[width=\textwidth]{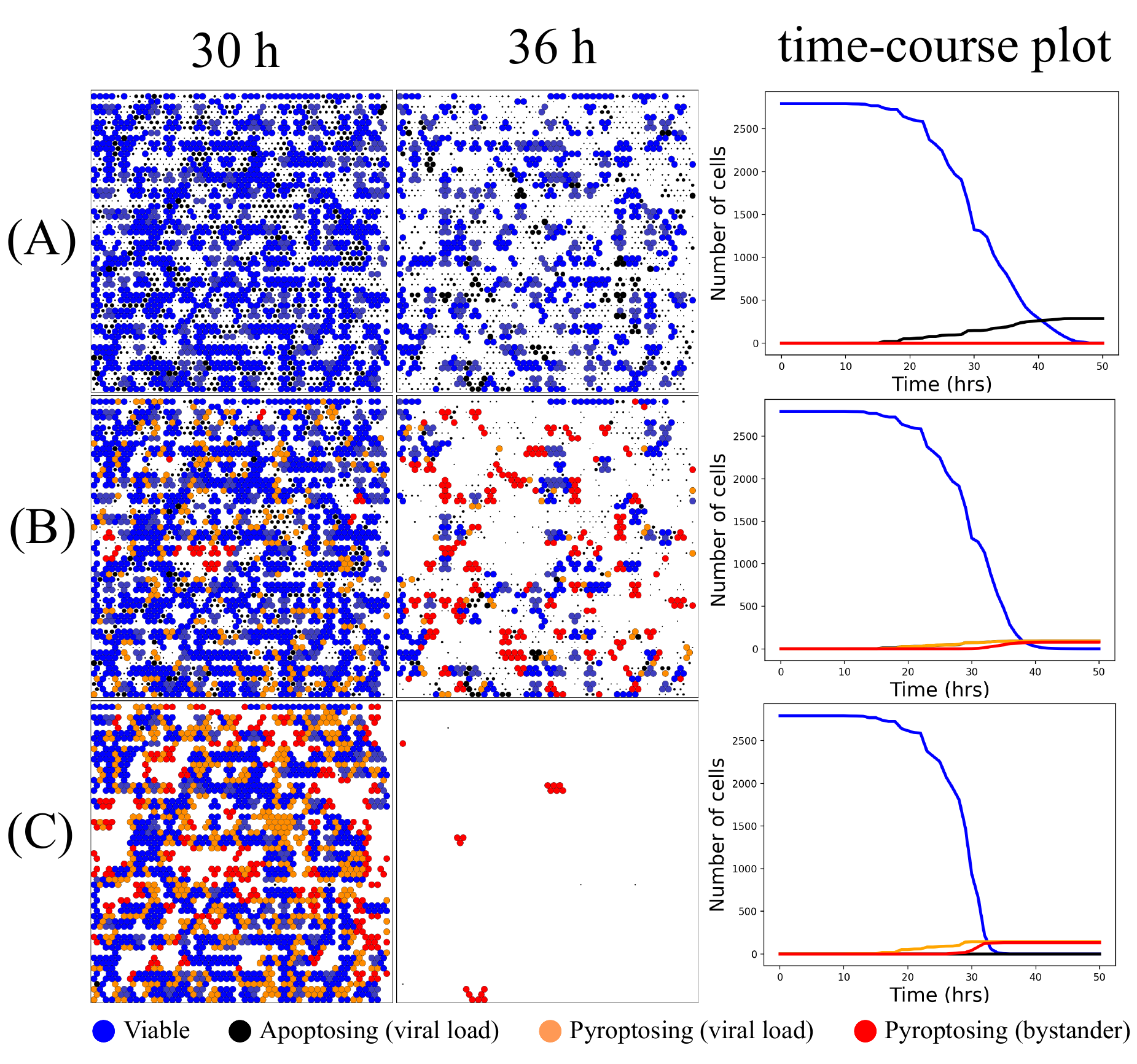}
 \caption{Cell map progressions and averaged time-course data for three situations, investigating the probability, $\Pi$, that a cell undergoes pyroptosis, instead of apoptosis, in response to viral load. We test three values: {\bf (A)} $\Pi=0$, {\bf (B)} $\Pi=0.5$ and {\bf (C)} $\Pi=1$. Viable cells (blue), infected cells (darker blue), cells undergoing apoptosis (black), cells undergoing viral induced pyroptosis (orange) and cells undergoing bystander induced pyroptosis (red) are shown in the cell maps, and the corresponding cell counts at each time-point are displayed in the time-course plots. Note, here $R_{\text{IL-1}\beta}=R_{\text{est}}$, the baseline value. }
 \label{fig:PhysiCell_Pi}
 \end{center}
\end{figure}

Simulation results are provided in form of cell maps and cell count data in Figures~\ref{fig:PhysiCell_IL1b} and \ref{fig:PhysiCell_Pi}. In the cell maps provided throughout this section, the results of a single run of the simulation are shown. Epithelial cells are shown in blue when they are viable, {\em i.e.,} neither pyroptosing nor apoptosing. 
Uninfected epithelial cells are shown in blue, whilst infected epithelial cells are depicted as darker blue proportional to their viral load. 
Cells in which the pyroptosis process has been initiated due to viral loads or bystander effects (initiated by internalised IL-1$\beta$ from the external environment) are respectively shown in orange and red. 
Once cell membrane rupture occurs, as determined by the intracellular pyroptosis pathway model, a pyroptosing cell is removed from the system. Apoptosing cells turn black and reduce in size until they are removed from the system.
The simulation data shown in the time-course plots is the total number of cells that are alive, and the accumulative number of cells that have initiated the investigated forms of cell death. In these time-course plots, the solid coloured lines denote the mean value of 10 runs of the simulation, and the standard deviation is plotted as a more transparent shade of the same colour. For all cases, the standard deviation between runs is generally small and thus, for clarity, we include a `zoomed-in' version of each of the time-course plots in Appendix~\ref{app:physicell}, as seen in Figure~\ref{fig:std10b}.
\\

\noindent We first study the impact of the bystander effect of IL-1$\beta$, where pyroptosis may be induced in infected or uninfected cells that are close to pyroptosing cells that are secreting the cytokine. 
In Figure~\ref{fig:PhysiCell_IL1b}, we display the results corresponding to three values of the threshold, $R_{\text{IL-1}\beta}$, specifically (A) $R_{\text{IL-1}\beta}=0.5\ R_{\text{est}}$, (B) $R_{\text{IL-1}\beta}= R_{\text{est}}$ and (C) $R_{\text{IL-1}\beta}=1.5\ R_{\text{est}}$, where {$R_{\text{est}}$} denotes an chosen threshold concentration. 
Comparing the three cell map progressions at the 30 h time-point, highlights qualitative differences between the results of the three cases. At this time-point, when we have a lower value of $R_{\text{IL-1}\beta}$, as seen in case (A), we observe a larger number of pyroptosing cells (mostly induced by the bystander effect) and fewer viable cells in the system, in comparison to the other cases, (B) and (C). In the median case (B), there is some bystander effect occurring, however not to as large an extent as case (A). When we have a higher value of $R_{\text{IL-1}\beta}$, as seen in case (C), we observe a lower bystander effect at the 30 h time-point. These qualitative differences are continued when comparing the 36 h time-point, whereby a lower value of $R_{\text{IL-1}\beta}$ (case (A)), results in all epithelial cells being removed from the system, while a higher value of $R_{\text{IL-1}\beta}$ (case (C)), results in a larger number of both viable and pyroptosing epithelial cells remaining. When comparing the time-course plots, we see that there is a significant difference in the number of cells that are induced to pyroptose through the bystander effect over time, where the bystander effect is the largest in case (A), and the smallest in case (C), as expected. We also observe that the time it takes for all viable cells to be removed from the system increases as we increase the threshold concentration required to induce pyroptosis through the bystander effect. 
These results highlight that the IL-1$\beta$ threshold concentration $R_{\text{IL-1}\beta}$ affects the ratio of the method of pyroptosis inducement (virus or IL-1$\beta$), and also the time it takes for all cells to be removed from the system, as pyroptosis is a more rapid mode of cell death than apoptosis. Although the results from these simulations are qualitative, the model provides a framework in which to study bystander effects. Upon availability, data related to the IL-1$\beta$ levels that are required for pyroptosis induction could be incorporated in the model for a more quantitative analysis. 
\\

\noindent We next study the impact of the probability that an infected cell undergoes pyroptosis, as opposed to apoptosis, in response to viral load, as determined by the parameter $\Pi$. 
In Figure~\ref{fig:PhysiCell_Pi} we display the results corresponding to three values of $\Pi$.
As before, comparing the three cell map progressions, the qualitative differences between the three cases are clear when comparing the cell maps at the 36 h time-point. In case (A) where the probability of pyroptosis is $\Pi=0$, the results show that a large number of viable epithelial cells remain in the system at 36 h. We observe that there are fewer viable cells and more pyroptosing cells in case (B) than in case (A) at 36 h, moreover the bystander effect is the main cause of pyroptosis in case (B). Finally, when $\Pi=1$ in case (C), by the 36 h time-point, all of the viable epithelial cells have been removed from the system, and only pyrotosing cells remain. 
Through the time-course plots, we can see a clear distinction between the number of cells undergoing apoptosis versus pyroptosis in the three cases. These simulation results show that increasing the value of $\Pi$ speeds up the overall eradication of epithelial cells. The reason for this is two-fold: pyroptosis contributes to bystander induced cell deaths, and the pyroptosis process is faster than apoptosis. More specifically, the time it takes for a cell to pyroptose or apoptose in the simulation is approximately 2 and 8.5 hours respectively. These results also highlight the impact of the bystander effect, which is, as expected, increased as $\Pi$ is increased, due to a larger number of viral load induced pyroptosing cells secreting IL-1$\beta$ into the external environment. As more cells are induced to pyroptose through the bystander effect, this further amplifies the levels of IL-1$\beta$ in the external environment, leading to an increased bystander effect. 
We additionally note, that in the time-course plot for case (B), where $\Pi=0.5$, we observe a higher standard deviation resulting from 10 simulation runs. This is due to the added stochasticity in the model arising from the non-deterministic mode of cell death when $\Pi \neq 0$ and $\Pi \neq 1$. The time-course plots for all cases are included in a `zoomed-in' format in Appendix~\ref{app:physicell}, Figure~\ref{fig:std10b}.

\FloatBarrier

\section{Conclusions and future work}
\label{sect:conclusions}
Pyroptosis has been identified as a key mechanism involved in the cytokine storm and the inflammation associated with severe cases of COVID-19.
Consequently, pyroptosis has been suggested as a target pathway to treat symptoms of COVID-19. 
In order to investigate the pathway of pyroptosis in further detail, we formulate a single-cell mathematical model that captures the key proteins and intracellular processes involved in this pro-inflammatory mode of cell death. 
Specifically, we model the process from DAMP/PAMP-induced TLR signalling to membrane rupture and the resulting cell death. 
The model is described in terms of a system of ODEs that captures the dynamics of intracellular molecule concentrations and cell volume. 
We further expand the model to include pharmacodynamic effects of a generic, NLRP3-targeting anti-inflammatory drug, and then perform both local and global sensitivity analyses in order to investigate how sensitive the model results are to the model parameters at their estimated values. 
Finally, we provide some detail on how this single-cell model is integrated into a wider cellular-scale model that is being used to investigate SARS-CoV-2 progression in lung epithelial cells~\cite{wang2020rapid}. 
\\

\noindent Recently, it has been found that if the pyroptosis pathway is inhibited, a cell can instead undergo apoptosis, a form of non-inflammatory cell death~\cite{Zheng2020,Kozloski2020,Taabazuing2017,Tsuchiya2020,he2015gasdermin,lee2018influenza}. 
In our single-cell model, described in Section~\ref{sect:model}, we have shown in our simulations that the inclusion of an NLRP3-targeting anti-inflammatory drug can delay the formation of the inflammasome, and therefore may provide a method of switching the cell death mode from inflammatory to non-inflammatory. 
The single-cell model developed in this study could further be extended to investigate more specific pharmacodynamic actions of anti-inflammatory drugs that target the formation of the NLRP3 inflammasome. A comprehensive review of such drugs is provided by Bertinara {\em et al.}~\cite{Bertinaria2019}. 
One such drug is tranilast, an NLRP3-targeting drug that is traditionally used for treating inflammatory diseases, such as asthma, in Japan and South Korea. Tranislast is currently being re-purposed for treating COVID-19 related symptoms in a clinical trial\footnote{\href{http://www.chictr.org.cn/showprojen.aspx?proj=49738}{Chinese Clinical Trial Registry, registration number:
ChiCDrug2000030002.}}~\cite{Yap2020,Lythgoe2020}. 
Moreover, corticosteroids such as dexamethasone have been suggested as agents to counter the cytokine storm in severe cases of COVID-19~\cite{sharun2020dexamethasone}, and preliminary reports from clinical trials suggest that low-dose dexamethasone therapy has beneficial effects in COVID-19 patients requiring ventilation or supplemental oxygen~\cite{recovery2020dexamethasone}.
As a corticosteroid, dexamethasone has many potential pharmacodynamic effects. 
Of relevance to the study of pyroptosis is that dexamethasone has been shown to directly bind to, and/or indirectly inhibit, NF-$\kappa$B, thus preventing the transcription of NLRP3 and pro-inflammatory cytokines~\cite{rhen2005antiinflammatory}. Therefore, in future work, we could mathematically investigate the impact that this type of drug action would have on the single-cell pathway model, and we could further investigate the interplay between apoptosis and pyroptosis.

When investigating the role of pyroptosis in the PhysiCell SARS-CoV-2 framework, as described in Section~\ref{sect:abm}, our results highlight that the mode of cell death impacts the cellular-scale dynamics. 
In this paper we have investigated an {\em in vitro} situation, however the PhysiCell framework supports the inclusion of within-host components, such as immune cells, to simulate {\em in vivo} scenarios. Previous studies have shown that IL-18 can act as a recruiter of immune cells~\cite{Bergsbaken2009,dinarello2009immunological,he2016mechanism,zalinger2017role,Stutz2009,oshea2015cytokines}. Therefore, processes that allow extracellular IL-18 to act as an immune-cell recruiting cytokine can be implemented in the PhysiCell SARS-CoV-2 framework, when simulating {\it in vivo} scenarios.

Our model currently does not consider the effects of negative-feedback within the pyroptosis pathway. However, we note that there are several recent experimental studies that highlight the potential existence of negative-feedback loops that regulate the pyroptosis pathway at a transcriptional level~\cite{cornut2020transcriptional,christgen2020toward}. For example, interferon regulatory factor 2 (IRF2) can bind to the pro-caspase-1 promoter, preventing pro-caspase-1 transcription, while IRF1 can upregulate this process~\cite{cornut2020transcriptional}. Additionally, melatonin has been shown to inhibit NLRP3 priming~\cite{chen2020melatonin,wu2020tlr2}. Our model could be adapted to include some of these regulatory mechanisms, to further investigate the single-cell level dynamics of pyroptosis. 

Investigating the role of interferons and interleukins within both the intracellular and cellular level would also be a beneficial extension to the model. At the intracellular level, it has been shown that interferon-1 (IFN-1) can decrease IL-18 transcription, and that interleukin-10 (IL-10) negatively regulates IL-1$\beta$ transcription~\cite{cornut2020transcriptional}. Thus, at the cellular level, IFNs can play a role in inhibiting pyroptosis initiation, reducing the bystander effect and preventing cytokine storms~\cite{Yap2020,mishra2013nitric,guarda2011type}. The role of type~I~IFNs in the treatment of COVID-19 is being investigated~\cite{lokugamage2020sars}, therefore it would be beneficial to investigate these mechanisms in both the pathway model and in the wider PhysiCell multiscale model, where other mechanisms of IFNs are currently being considered.
\\

\noindent We finally remark that pyroptosis has been identified as a potential contributor to symptoms in various diseases other than COVID-19~\cite{Yap2020}, such as other virus-induced diseases~\cite{Stutz2009}, auto-immune diseases~\cite{Bertinaria2019,Bergsbaken2009} and cancer~\cite{Zheng2020,Fink2005,lee2018influenza}. Therefore, the mathematical pyroptosis model developed in this paper could be modified and used to investigate the effect of pyroptosis in other disease scenarios.

\section*{Acknowledgments}
\noindent SJH is supported by the Medical Research Council grant MR/R017506/1.
\noindent The authors would like to thank the SARS-CoV-2 Tissue Simulation Coalition for ongoing collaboration and feedback on model development.
Further, this work is part of the RAMP (Rapid Assistance in Modelling the Pandemic) initiative, coordinated by the Royal Society, UK. The authors would like to thank Prof. Mark A.J. Chaplain (University of St Andrews) for coordinating our RAMP Task Team modelling within-host dynamics.
\\

\noindent SARS-CoV-2 Tissue Simulation Coalition: \url{http://physicell.org/covid19/}.

\noindent RAMP: \url{https://royalsociety.org/topics-policy/health-and-wellbeing/ramp/}.

\newpage
\renewcommand\thefigure{\thesection.\arabic{figure}} 
\renewcommand{\thetable}{\thesection.\arabic{table}}
\appendix
\setcounter{figure}{0} 
\setcounter{table}{0}

\section{Abbreviations used in text}
\label{appendix_abb_intext}
\begin{itemize}
 \item ARDS: acute respiratory distress syndrome 
 \item ASC: apoptosis-associated speck-like protein containing a CARD
 \item CARD: caspase activation and recruitment domain 
 \item COVID-19: coronavirus disease 2019
 \item CD: Catalytic domain
 \item C1-CD: the catalytic domain of caspase-1
 \item DAMP: damage associated molecular pattern
 \item GSDMD: gasdermin D
 \item GSDMD-N: N-terminal domain of GSDMD
 \item IFN: interferon
 \item IL-1$\beta$: interleukin 1$\beta$
 \item IL-18: interleukin 18
 \item LLR: leucine-rich repeat
 \item NBD: nucleotide-binding and oligomerisation domain
 \item NF-$\kappa$B: nuclear factor kappa-light-chain-enhancer of activated B cells
 \item NLRP3: NBD, LRR-containing receptors with an N-terminal PYD 3
 \item ODE: ordinary differential equation
 \item PAMP: pathogen associated molecular pattern
 \item pro-IL-1$\beta$: pro-form of interleukin 1$\beta$
 \item pro-IL-18: pro-form of interleukin 18
 \item PYD: pyrin domain
 \item SARS-CoV-2: severe acute respiratory syndrome coronavirus 2
 \item TLR: toll like receptor
\end{itemize}

\section{Detailed model description and motivation}
\label{app_description}

In this appendix we provide further details of the modelling terms chosen within our single-cell pathway model described in Section~\ref{sect:model}. 
Model variables/components are listed in Table~\ref{tab:var} and the model parameters are discussed {in Appendix~\ref{app_parameters}}.
\begin{table}
\centering
\begin{tabular}{ |l|l|l| } 
 \hline
 Label & Symbol & Component Description \\
 \hline
 $y_1$&$\text{[NF-$\kappa$B\textsubscript{n}]}(t)$ & concentration of nuclear NF-$\kappa$B \\
 $y_2$& $\text{[NLRP3\textsubscript{i}]}(t)$ & concentration of inactive NLRP3 protein\\
 $y_3$&$\text{[NLRP3\textsubscript{a}]}(t)$ & concentration of active NLRP3 protein\\
 $y_4$&$\text{[NLRP3\textsubscript{o}]}(t)$ & concentration of oligomerised NLRP3 protein\\
 $y_5$&$[\text{ASC\textsubscript{b}}](t)$ & concentration of bound ASC protein\\
 $y_6$&$\text{[C1]}(t)$ & concentration of activated caspase-1\\
 $y_{7}$&$\text{[GSDMD-N]}(t)$ & concentration of activated GSDMD-N\\
 $y_{8}$&$[\text{pro-IL-1$\beta$}](t)$ & concentration of pro-IL-1$\beta$\\
 $y_{9}$&$\text{[IL-1$\beta$\textsubscript{c}]}(t)$ & concentration of active IL-1$\beta$ in the cytoplasm\\
 $y_{10}$&$\text{[IL-1$\beta$\textsubscript{e}]}(t)$ & concentration of active IL-1$\beta$ external\\
 $y_{11}$&$[\text{IL-18}_{\text{c}}](t)$ & concentration of active IL-18 in the cytoplasm\\
$y_{12}$&$[\text{IL-18}_{\text{e}}](t)$ & concentration of active IL-18 external\\
 $y_{13}$&$[\text{Drug}](t)$ & concentration of free anti-inflammatory drug \\
 $y_{14}$&$[\text{Drug} \cdot \text{NLRP3}_{\text{a}}](t)$ & concentration of anti-inflammatory drug - NLRP3$_{\text{a}}$ complex \\
 $y_{15}$&$V(t)$ & cell volume \\
 \hline
\end{tabular}
\vspace{0.1in}
 \caption{The symbols and descriptions of the components considered in the model. The bracket notation [ ] denotes compound concentration. Implicit time-dependence is denoted by $(t)$.} \label{tab:var}
\end{table}

\subsection{Signal 1.}
The initial signal, $S_{1}$, as defined in Equation~\eqref{eqS1}, is here considered to be either {\it on}, $S_1=1$, or {\it off}, $S_1=0$. This binary signal is motivated by the modelling assumption that there will be no inflammasome-inducing signalling until the DAMP/PAMP-sensing TLR activity is high enough. Therefore any signal level below some threshold value, here \emph{e.g.,} $S_{1}=1$, would not stimulate a downstream response. 
We assume that once $S_1$ is turned {\it on}, this signal is kept at a constant level until the cell dies. 

\subsection{Signal 2.}
In the model, we assume that the second signal, $S_{2}$, is either {\it on} or {\it off}. For all shown simulation results, we assume that $S_2$ is {\it on}, so that $S_{2}=1$, for all time-points in the simulation. Through experiments, Juliana {\em et al.}~\cite{juliana2012non} showed that increasing the time that a cell was exposed to ATP (yielding $S_2$ to turn {\it on}) did not alter the subsequent inflammasome dynamics, nor the expression of involved proteins. 
Thus, assuming that $S_2$ is binary, {\em i.e.}, {\it on} or {\it off}, throughout the simulation is a reasonable model assumption. 

\subsection{Reaction terms.}
\label{sect:reaction_terms}
To describe the transcription or cleavage terms we use a Hill function~\cite{Salahudeen_ligandbinding} of the form, 
\begin{equation}
 \alpha \frac{[A]^{\gamma}}{A_{50}^{\gamma}+[A]^{\gamma}}.\label{Hill2_int} 
\end{equation}
Here, A\textsubscript{50} represents the level of [A] required for the Hill function to reach the value of a half, often referred to the half-max value of [A]~\cite{lolkema2015hill}. The rate parameter $\alpha$ determines the cleavage or transcription rate. 
To describe the activation terms in the ODE system we consider chemical reactions of the form,
\begin{equation}
A \xrightarrow{k} B,\label{activation}
\end{equation}
where $A$ and $B$ are the inactive and active components, respectively. The reaction rate $k$ determines how fast this process occurs. Using the law of mass action we can then translate this into the term used in the ODE for concentration $[A]$, {\em i.e.,} $- k[A]$, and for concentration $[B]$, {\em i.e.,} $+ k[A]$.

Similarly, to include decay terms, we consider a chemical reaction of the form,
\begin{equation}
A \xrightarrow{\delta}\emptyset,\label{decay}
\end{equation}
where $\emptyset$ here denotes that the component is removed from the system. Again, using the law of mass action we can then use this to write the decay terms within the ODE for concentration $[A]$, {\em i.e.,} $- \delta [A]$.

Catalytic interactions within the model are added considering a reaction of the form,
\begin{equation}
A+B\xrightarrow{k} C+B.\label{binding}
\end{equation}
Here, component $A$ and $B$ bind together briefly to transform $A$ into a 3rd component $C$. Using the law of mass action, we can then use this to write the reactions within the ODE for concentration $[A]$, {\em i.e.,} $- k [A][B]$, and concentration $[C]$, {\em i.e.,} $+k [A][B]$. Note, as $B$ is not used up the terms for concentration $[B]$ will will not be included.

We consider the NLRP3 oligomerisation process to be accumulative, that is once two active NLRP3 proteins bind, then any additional proteins will bind on to this bound NLRP3 forming the inflammasome. If p proteins are required to form the inflammasome the interactions will be of the form,
$$
\text{[NLRP3\textsubscript{a}]}+\text{[NLRP3\textsubscript{m-1}]} \xrightarrow{k_{2}} \text{[NLRP3\textsubscript{m}]} \quad \mbox{for}\ \text{m}=2,\ldots,\text{p},
$$ 
where $m$ is the number proteins bound together. Note, we have that [NLRP3\textsubscript{1}]=[NLRP3\textsubscript{a}] as this is unbound active protein.

The law of mass action gives for m=2,$\ldots$,p-1,
\begin{eqnarray}
\frac{d \text{[NLRP3\textsubscript{m}]}}{dt}&=&k_{2} \text{[NLRP3\textsubscript{a}]}\text{[NLRP3\textsubscript{m-1}]}-k_{2} \text{[NLRP3\textsubscript{a}]}\text{[NLRP3\textsubscript{m}]}\nonumber\\
&=& k_{2} \text{[NLRP3\textsubscript{a}]} \left(\text{[NLRP3\textsubscript{m-1}]}-\text{[NLRP3\textsubscript{m}]}\right),\nonumber
\end{eqnarray}
and
$$
\frac{d \text{[NLRP3\textsubscript{p}]}}{dt}=k_{2} \text{[NLRP3\textsubscript{a}]}\text{[NLRP3\textsubscript{p-1}]}.
$$

The total level of bound or oligomerised protein is,
$$
\text{[NLRP3\textsubscript{o}]}=\sum_{\text{m}=2}^{\text{p}} \text{[NLRP3\textsubscript{m}]}.
$$
Considering the rate of change of [NLRP3\textsubscript{o}] we obtain, 
\begin{eqnarray}
&&\frac{d\text{[NLRP3\textsubscript{o}]}}{dt}=\sum_{\text{m}=2}^{\text{p}} \frac{d\text{[NLRP3\textsubscript{m}]}}{dt}\nonumber\\
&&=k_{2} \text{[NLRP3\textsubscript{a}]} \left( \text{[NLRP3\textsubscript{a}]}-\text{[NLRP3\textsubscript{2}]}+\sum_{\text{m}=3}^{\text{p}-1} \left(\text{[NLRP3\textsubscript{m-1}]}-\text{[NLRP3\textsubscript{m}]}\right) + \text{[NLRP3\textsubscript{p-1}]}\right).\nonumber
\end{eqnarray}
This simplifies to,
$$
\frac{d\text{[NLRP3\textsubscript{o}]}}{dt}=k_{2} \text{[NLRP3\textsubscript{a}]}^2,
$$
as all other terms cancel out.

\subsection{Anti-inflammatory drug terms.}
The influence of the anti-inflammatory drug is included in the modified version of the model which includes additional terms for the dynamics of NLRP3\textsubscript{a}, as given in Equation~\eqref{eqNa_drug}. Also included in this modified model are the dynamics of the anti-inflammatory drug, Drug, and the complex Drug~$\cdot$~NLRP3\textsubscript{a}, as listed in Equations~\eqref{eqDrug_1b} and~\eqref{eqDrug_1}. Here, we consider a binding reaction of the form,
\begin{equation}
\text{Drug + NLRP3\textsubscript{a}} \underset{k_{\text{-D}}}{\stackrel{k_{\text{+D}}}{\rightleftharpoons}} \text{Drug} \cdot \text{NLRP3\textsubscript{a}}.\label{Drug_int}
\end{equation}
Mass action kinetics allows us to write this interaction in terms of ODEs. We consider the anti-inflammatory drug to be a covalent inhibitor, and thus we could add an extra step in the above reaction. However, in lieu of relevant data, we are currently considering only one reaction step for Drug binding to NLRP3\textsubscript{o}, in an effort to minimise the model.

\setcounter{figure}{0} 
\setcounter{table}{0}
\renewcommand{\thetable}{C.\arabic{table}}
\renewcommand\thefigure{C.\arabic{figure}}
\renewcommand{\theHtable}{Supplement.\thetable}
\renewcommand{\theHfigure}{Supplement.\thefigure}
\section{Model parameter values}
\label{app_parameters}
In this appendix, we provide further details on how the model parameter values, listed in Table~\ref{tab:par}, were estimated. 
\begin{table}
\centering
\resizebox{1.1\textwidth}{!}{\begin{tabular}{ |l|l|l|l| } 
 \hline
 Label & Symbol & Description & Value \\
 \hline
 & $a$ & sigmoid constant for step-function approximation & 1 (a.u.) \\
$\theta_{1}$&$\alpha_{1}$ & transcription rate of NLRP3\textsubscript{i} & 0.07 (a.u.)\ min$^{-1}$ \\
$\theta_{2}$&$\alpha_{2}$ & cleavage rate of GDSMD & 0.1 min$^{-1}$ \\
$\theta_{3}$&$\alpha_{3}$ & transcription rate of pro-IL-1$\beta$ & 0.06 (a.u.)\ min$^{-1}$ \\
$\theta_{4}$&$\alpha_{4}$ & cleavage rate of pro-IL-1$\beta$ & 1 min$^{-1}$ \\
$\theta_{5}$&$\alpha_{5}$ & cleavage rate of pro-IL-18 & 1 min$^{-1}$ \\
 &$b$ & sigmoid constant for step-function approximation & 2 (a.u.) \\
 &$c$ & sigmoid constant for step-function approximation & 1000 \\
$\theta_{6}$&C1\textsubscript{50} & half-max value of caspase-1 required for cleavage & 0.3 a.u. \\
$\theta_{7}$&$\delta_{1}$ & decay rate of NLRP3\textsubscript{i} and NLRP3\textsubscript{a} & 0.002 min$^{-1}$ \\
$\theta_{8}$&$\delta_{2}$ & decay rate of pro-IL-1$\beta$ and IL-1$\beta\textsubscript{c}$& 0.004 min$^{-1}$ \\
$\theta_{9}$&$h$ & maximum height elevation of the nuclear NF-$\kappa$B peak & 0.55 a.u. \\
$\theta_{10}$&$\gamma_{\text{C1}}$ & Hill function coefficient of caspase-1 mediated cleavage & 2 \\
$\theta_{11}$&$\gamma_{\text{NF}}$ & Hill function coefficient of nuclear NF-$\kappa$B mediated transcription & 2 \\
$\theta_{12}$&$k_{1}$ & activation rate of NLRP3\textsubscript{i} & 0.7 min$^{-1}$ \\
$\theta_{13}$&$k_{2}$ & oligomerisation rate of NLRP3\textsubscript{a} &1 (a.u.)$^{-1}$min$^{-1}$ \\
$\theta_{14}$&$k_{3}$ & rate for NLRP3\textsubscript{a} - ASC\textsubscript{f} interaction
&0.04 (a.u.)$^{-1}$\ min$^{-1}$ \\
$\theta_{15}$&$k_{4}$ & rate for ASC\textsubscript{b} - (pro-caspase-1) interactions &0.03 (a.u.)$^{-1}$\ min$^{-1}$ \\
$\theta_{16}$&$k_{5}$ & transport rate of IL-1$\beta$\textsubscript{c} out of cell & 1 min$^{-1}$ \\
$\theta_{17}$&$k_{6}$ & transport rate of IL-18\textsubscript{c} out of cell & 1 min$^{-1}$ \\
$\theta_{18}$&$k_{7}$ & rate at which cell volume increases & 0.2 min$^{-1}$ \\
&$k_{+\text{D}}$ & forward rate constant for the drug-NLRP3\textsubscript{a} reaction& 0.005 (a.u.)$^{-1}$\ min$^{-1}$ \\
&$k_{-\text{D}}$ & reverse rate constant for the drug-NLRP3\textsubscript{a} reaction & 0.00005 min$^{-1}$ \\
&$n$& level of NLRP3$_{\text{o}}$ required to form the inflammasome base & 1 a.u. \\
$\theta_{19}$&NF\textsubscript{50} & half-max value of nuclear NF-$\kappa$B required for transcription & 0.3 a.u. \\
$\theta_{20}$&$s$ & skewness of the nuclear NF-$\kappa$B peak & 0.8 \\
$\theta_{21}$&$\tau$ & time when the nuclear NF-$\kappa$B peak occurs & 10 min \\
 & $V_{c}$ & critical volume at which cell ruptures & 1.5 a.u. \\
 \hline
\end{tabular}}
\vspace{0.1in}
 \caption{Parameters values used within numerical simulations of the model. The justification for these values is given in Appendix~\ref{app_parameters}. The parameters investigated through sensitivity analysis are assigned a label, $\theta_{n}$. Here `a.u.' refers to arbitrary units of concentration or volume. 
 \label{tab:par} }
\end{table}

\subsection{Decay parameters.}
In~\cite{han2015lipopolysaccharide}, it was found that NLRP3 had a half-life of approximately 6 hrs when cells are stimulated with lipopolysaccharide (LPS), which translates to a decay rate of approximately 0.002 $\mu\mbox{m} \mbox{ min}^{-1}$. Therefore we set,
$$
\delta_{1}= 0.002\ \text{min}^{-1},
$$
as we assume measure concentrations as arbitrary units. In~\cite{moors2000proteasome}, it was found that pro-IL-1$\beta$ had a half-life of approximately 2.5 hrs in human monocytes, which translates to a decay rate of approximately 0.004 $\mu\mbox{m}\ \mbox{min}^{-1}$. Therefore we set,
$$
\delta_{2}= 0.004\ \text{min}^{-1},
$$
as we assume measure concentrations as arbitrary units.
\begin{figure}
 \centering
 \includegraphics[scale=0.9]{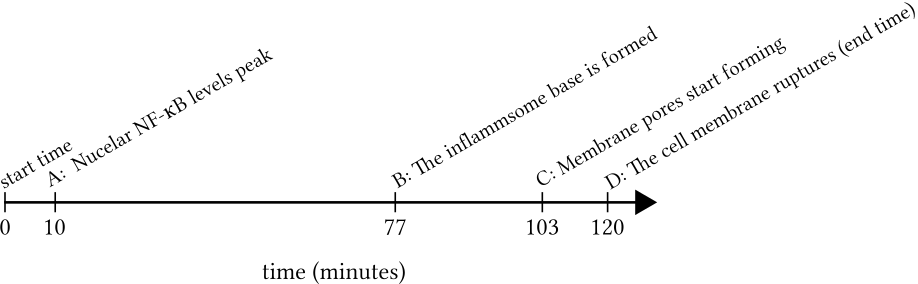}
 \caption{We construct a timeline for the pyroptosis pathway using data from Bagaev {\em et al.}~\cite{bagaev2019elevated}, de Vasconcelos {\em et al.}~\cite{deVasconcelos2019} and Mart{\'\i}n-S{\'a}nchez {\em et al.}~\cite{martin2016inflammasome}. 
 At 0 minutes (start time), Signal 1 ($S_{1}$) turns {\it on}. Thereafter, at 10 minutes, nuclear NF-$\kappa$B levels peak, after which the cytoplasmic-to-nuclear translocation of NF-$\kappa$B ceases in the mathematical model. 77 minutes into the timeline, enough NLRP3 has been transcribed, synthesised and activated to form the inflammasome base. Then, at 107 minutes, the GSDMD-N induced pores start forming, allowing the outflux of IL-1$\beta$ and IL-18, as well as the influx of extracellular water which causes the cell volume to start increasing. Finally, at 120 minutes (the end time) the cell membrane completely ruptures. At this point, around 90\% of the pro-IL-1$\beta$ and pro-IL-18 will have been cleaved and subsequently released to the extracellular environment.
 }
 \label{fig:pyro_timeline}
\end{figure}

\subsection{The timeline of pyroptosis.}
We use data from three experimental studies~\cite{deVasconcelos2019, bagaev2019elevated,martin2016inflammasome} to estimate a timeline for events regarded in our mathematical model. 
The values of the model parameters $\alpha_{1}-\alpha_{5},\ \text{C1}_{50},\ \gamma_{\text{C1}},\ \gamma_{\text{NF}},\ k_{1}-k_{7}, \ \text{NF}_{50}$ and $V_{c}$ are chosen to obtain model dynamics that follow this experimentally motivated timeline of events \textbf{A-D}, which is illustrated in Figure \ref{fig:pyro_timeline}. 
Further detail of constructing this timeline is provided in the following paragraphs.

Bagaev {\em et al.}~\cite{bagaev2019elevated}
studied NF-$\kappa$B kinetics in primary macrophages subjected to bacterial lipopolysaccharide (LPS). de Vasconcelos {\em et al.}~\cite{deVasconcelos2019} investigated subcellular key events that define pyroptosis using single-cell analysis. In their article, the authors provide a multitude of data pertaining to bone marrow-derived macrophages (BMDMs) that undergo caspase-1-dependent pyroptosis upon Bacillus anthracis lethal toxin (LeTx) stimulation. Mart{\'\i}n-S{\'a}nchez {\em et al.}~\cite{martin2016inflammasome} measured the levels of pro-IL-1$\beta$ and external IL-1$\beta$ over time after inflammasome activation in systems with mouse BMDMs.\\

\label{par:data}
\textbf{A}: Bagaev {\em et al.}~\cite{bagaev2019elevated} reported that the nuclear NF-$\kappa$B concentration peaked at 10 minutes post LPS activation, after which it decreased to a half-maximal level in a gradual manner over 100 minutes.\\

\textbf{B}: de Vasconcelos {\em et al.}~\cite{deVasconcelos2019} showed that the influx of Ca$^{2+}$ occurs before total membrane permeabilisation in pyroptosis (see Figure 7 in de Vasconcelos {\em et al.}~\cite{deVasconcelos2019}). Further data from their study suggests that changes in mitochondrial activity occur between (45+12=56) and (21+9=30) minutes prior to complete membrane rupture (see Figure 1 combined with Figure 3, as well as Supplementary Figure 3 in de Vasconcelos {\em et al.}~\cite{deVasconcelos2019}). If we assume that Ca$^{2+}$ influx and changes in mitochondrial activity commence between the priming and activation events, we can estimate that the NLRP3 inflammasome base is formed somewhere between 56 to 30 minutes prior to cell rupture. In the model we use the average value (43 minutes) as the time before cell rupture (end time) when the inflammasome forms ({\em i.e.} when the concentration of NLRP3\textsubscript{o} reaches the threshold value $n$). Therefore event \textbf{B} occurs around end time - 43 minutes = (120-43) minutes = 77 minutes.\\

\textbf{C}: In de Vasconcelos {\em et al.}'s study, cell volumes increase gradually for approximately 13 minutes prior to membrane rupture. Furthermore, cell volumes were reported to increase by approximately 50\%. (See Figure 6a,b in de Vasconcelos {\em et al.}~\cite{deVasconcelos2019}).\\

\textbf{D}: Supernatants from cell cultures were assayed for LDH activity in de Vasconcelos {\em et al.}'s study. LDH is a cytosolic protein that is released by pyroptotic cells, and LDH activity is commonly used as a marker for plasma rupture in experimental settings. Data showed that near-maximal supernatant LDH values were reached 120 minutes post LeTx stimulation (see Figure 9c in de Vasconcelos {\em et al.}~\cite{deVasconcelos2019}). From this we approximate the time between signal 1 turning {\it on} in the model, and complete membrane rupture, to be approximately 120 minutes.\\

Furthermore, results from Mart{\'\i}n-S{\'a}nchez {\em et al.}'s~\cite{martin2016inflammasome} study highlight that the release of IL-1$\beta$ coincided with membrane permeability (the formation of GSDMD-N derived membrane pores), and that eventually all of the pro-IL-1$\beta$ present at the start of the experiments was cleaved and released from the cell, with approximately 90\% released within 120 minutes.

\subsection{Anti-inflammatory drug parameters.}
We consider drug concentrations to be constant during the simulated time span, effectively describing an {\it in vitro} scenario without drug wash out. This could be altered to include drug dosages varying in time if appropriate. Tranilast and oridonin examples of anti-inflammatory drugs that can inhibit the formation of the NLRP3 inflammasome by binding to active NLRP3 \cite{Bertinaria2019}. 
Immunoblot data presented by Huang {\em et al.}~\cite{Huang2018}, pertaining to LPS-primed bone marrow derived macrophages (BMDMs) treated with tranilast, demonstrate dose-dependent (25-100$\mu$M) drug responses, where the maximal inhibition was reached at at 100 $\mu$M of the drug. 
In our model, we implicitly assume that pyroptosis is inhibited if the membrane rupture does not occur within 300 simulated minutes. 
We here set $k_{-\text{D}}/k_{+\text{D}}=0.01 \text{ a.u.}$, so that pyroptosis is inhibited for drug concentrations over 0.75 a.u..

\section{ {\sc \bf{MATLAB}} code and set up of numerical simulations for the single-cell model}
\label{app:code_and_numerics}
The single-cell mathematical model was implemented in {\sc MATLAB} using the ODE solver \texttt{ode15s}. 
The simulation results provided in Figures \ref{fig:basecase} and \ref{fig:completedrug} are obtained by running the code with the initial conditions given in Section~\ref{sect:implement}, along with the parameters given in Table~\ref{tab:par} and the signals $S_1$ and $S_2$ turned {\it on}. The {\sc MATLAB} code used to run these simulations is available on our project GitHub page: \url{https://github.com/Pyroptosis}.

\setcounter{figure}{0} 
\setcounter{table}{0}
\renewcommand{\thetable}{E.\arabic{table}}
\renewcommand\thefigure{E.\arabic{figure}}
\renewcommand{\theHtable}{Supplement.\thetable}
\renewcommand{\theHfigure}{Supplement.\thefigure}
\section{Details of the PhysiCell implementation}
\label{app:physicell}

\begin{figure}[H]
\begin{center}
 \includegraphics[width=\textwidth]{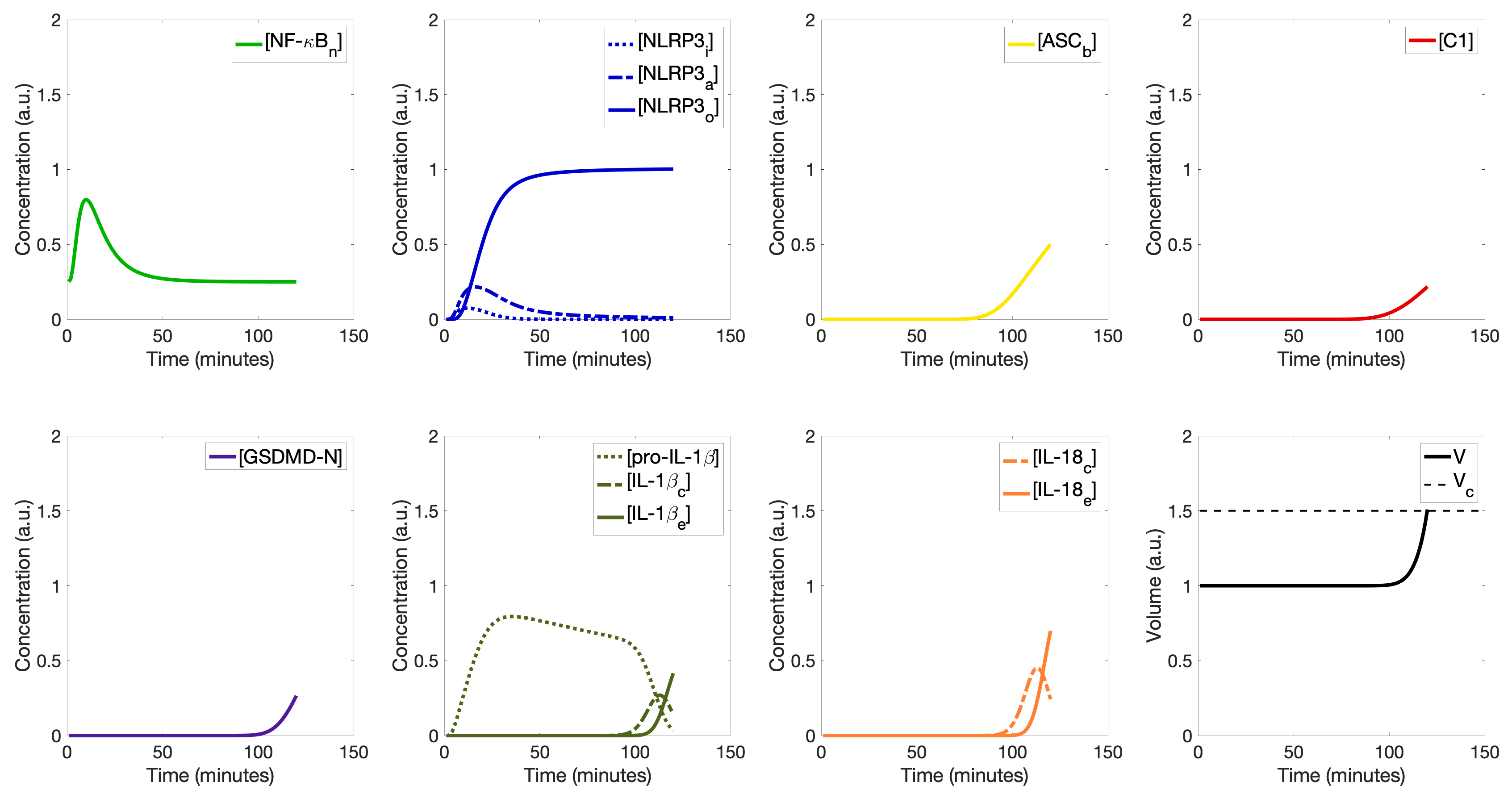}
 \caption{ Numerical simulations results of the Backward Euler discretisation of the ODE model described by system~\eqref{full_system} with functions~\eqref{full_functions}. The parameters are displayed in Table~\ref{tab:par} and here we use a time-step of $\Delta_t=1$ min. We display the concentration of each model component in arbitrary units (a.u.) over time, as well as cell volume dynamics. Note that the time progression stops once the critical volume $V_c$ is reached and the ultimate cell membrane rupture occurs.}
 \label{fig:backward}
 \end{center}
\end{figure}

To implement the ODE pathway model, as described in Section~\ref{sect:model}, into the PhysiCell framework. We discretise the full system~\eqref{full_system} with functions~\eqref{full_functions} using the Backward Euler method to implement it in to the PhysiCell framework. Note, that we do not consider the drug in the PhysiCell setting, that is we let $[\text{Drug}](t)\equiv 0$. The parameter values used are those previously shown in Table~\ref{tab:par}. The base threshold level of IL-1$\beta$ required to initiate pyroptosis is denoted by parameter $R_{\text{est}}$ and the threshold level of viral RNA required to initiate viral death is denoted by $R_{\text{RNA}}$. For the results shown in Figures~\ref{fig:PhysiCell_IL1b}~and~\ref{fig:PhysiCell_Pi}, we set,
\begin{equation}
R_{\text{est}}=200 \text{ a.u.} \quad \text{and} \quad R_{\text{RNA}}= 2 \text{ RNA}.
\end{equation}
The values of the parameters $\Pi$ and $R_{\text{IL-1}\beta}$ are varies, and shown in the caption of each Figure. We additionally plot the solutions of the Backward Euler method using these parameters in Figure~\ref{fig:backward}. For more details of the full PhysiCell model, we refer the reader to the most recent preprint~\cite{wang2020rapid}, or the SARS-CoV-2 Tissue Simulation Coalition webpage: \url{http://physicell.org/covid19/}. 
Additionally, the {\sc MATLAB} code used to create Figure~\ref{fig:backward}, and the specific PhysiCell C\texttt{++} code used to create Figures~\ref{fig:PhysiCell_IL1b}~and~\ref{fig:PhysiCell_Pi} is available on our project GitHub page: \url{https://github.com/Pyroptosis}.

\subsection{Apoptosis model in PhysiCell}
The following details are taken from~\cite{physicellsimulator}, and describes the in built standard apoptosis model in PhysiCell. Cell volume is tracked through multiple properties including,
\begin{itemize}
 \item $V_F$ - total fluid volume of the cell,
 \item $V_{CS}$ - total solid cytoplasmic volume of the cell,
 \item $V_{NS}$ - total solid nuclear volume of the cell.
\end{itemize}
Once apoptosis initiates, the cell volume changes according to the following equations,
\begin{eqnarray}
\frac{d V_F}{dt}&=& -r_F V_F,\\
\frac{d V_{CS}}{dt}&=& -r_{CS} V_{CS},\\
\frac{d V_{NS}}{dt}&=& -r_{NS} V_{NS},
\end{eqnarray}
where the parameters $r_F,\ r_{CS},\ r_{NS}$ are the rate of change of the total fluid volume, the solid cytoplasmic volume and the solid nuclear volume, respectively.

In the PhysiCell model, the cell is removed once a threshold volume $V_m$ has been reached, or the cell has been apoptosing for longer than time $T_A$. The default parameter values for apoptosis of epithelial cells~\cite{physicellsimulator} are,
$$
r_F = 0.05 \ \text{min}^{-1}, \quad r_{CS} = \frac{1}{60} \ \text{min}^{-1}, \quad r_{NS} = \frac{0.35}{60} \ \text{min}^{-1},
$$
$$
 T_A=8.6\ \text{hrs} = 516\ \text{mins}, \quad V_m=2\ \mu\text{m}^{3} .
$$

We consider that along with natural cell death, that the virus can induce apoptosis or pyroptosis once viral RNA has reached a certain threshold, $R_{\text{RNA}}$.

\newpage
\begin{figure}[H]
\begin{center}
 \includegraphics[width=\textwidth]{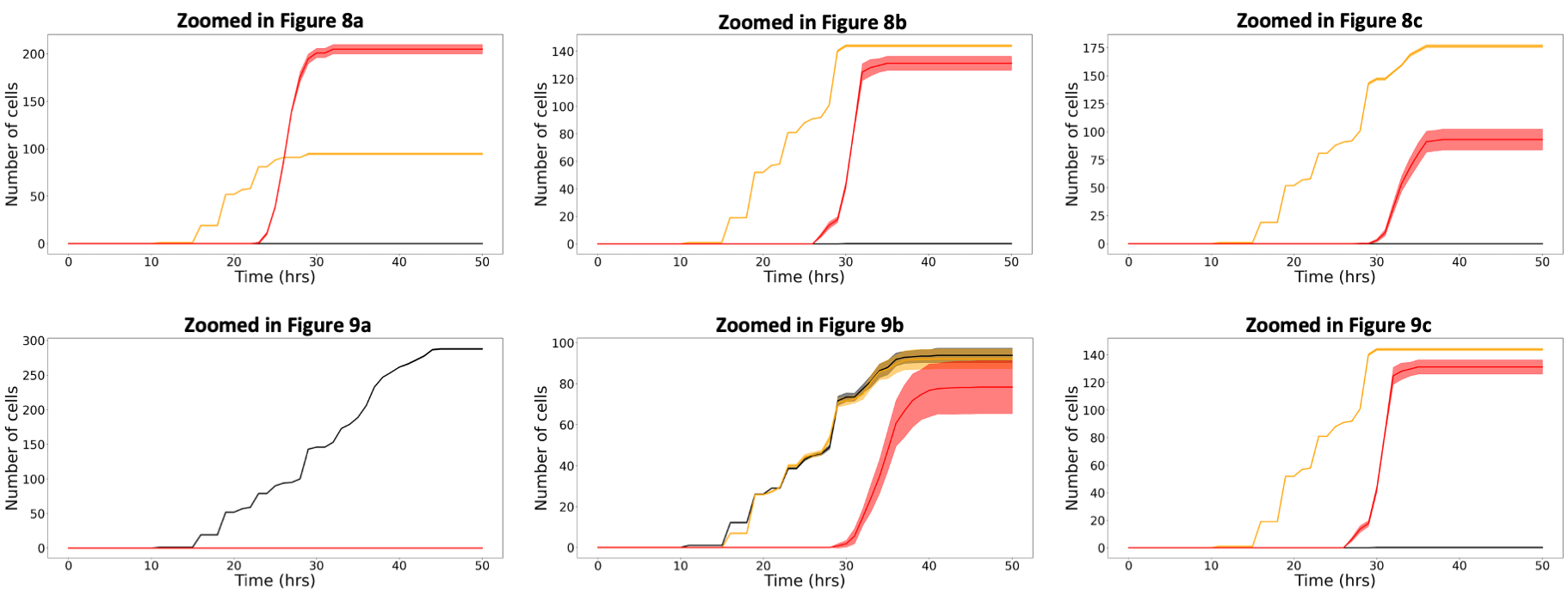}
 \caption{ The averaged time-course data for all cases from Figure~\ref{fig:PhysiCell_IL1b} and Figure~\ref{fig:PhysiCell_Pi} with viable cells omitted. Cells undergoing apoptosis (black), cells undergoing viral induced pyroptosis (orange) and cells undergoing bystander induced pyroptosis (red) are shown in the plots. The average over 10 runs is plotted as an opaque line, where the transparent colour represents the standard deviation. }
 \label{fig:std10b}
 \end{center}
\end{figure}

\end{document}